\documentclass[%
 reprint,
 amsmath,amssymb,
 aps,
prb,
]{revtex4-1}
	\usepackage{feynmf}
 \usepackage{tabularx}
	\usepackage{graphics}
	\usepackage{epsfig}
	\usepackage{graphicx}
	\usepackage{color}
	\usepackage{comment}
 \usepackage[scr=boondoxo]{mathalpha}

\begin{document}
	\unitlength = 1mm

\title{Spin-Phonon Coupling in Transition Metal Insulators: General Computational Approach and Application to MnPSe$_3$}

\author{Ramesh Dhakal}
\affiliation{Department of Physics and Center for Functional Materials, Wake Forest University, Winston-Salem, North Carolina 27109, USA}

\author{Samuel Griffith}
\affiliation{Department of Physics and Center for Functional Materials, Wake Forest University, Winston-Salem, North Carolina 27109, USA}

\author{Kate Choi}
\affiliation{Department of Physics and Center for Functional Materials, Wake Forest University, Winston-Salem, North Carolina 27109, USA}

\author{Stephen M. Winter}
\email{winters@wfu.edu}
\affiliation{Department of Physics and Center for Functional Materials, Wake Forest University, Winston-Salem, North Carolina 27109, USA}
\date{\today}

\begin{abstract}
Spin-phonon coupling underlies a number of diverse range of phenomena of recent interest, particularly in transition metal insulators with strong spin-orbit effects, where it may give rise to hybrid magnetoelastic excitations, and the controversial phonon thermal Hall effect. In this work, we describe a general approach to the first principles estimation of generic spin-orbit-phonon couplings suitable for investigating these diverse effects. The method is demonstrated by application to study the temperature-dependent optical phonon lineshapes in MnPSe$_3$, for which we find quantitative agreement with recent experiments.
\end{abstract}

\maketitle

\section{Introduction}

Spin-lattice coupling underlies a wide variety of phenomena in magnetic transition metal insulators. For example, it is considered a key ingredient for optimization of relaxation times in spin-qubits\cite{lunghi2022toward,kragskow2023spin}, the thermal transport properties of magnon spintronics\cite{bozhko2020magnon}, and the properties of 2D layered magnets pursued for magnetic heterostructure devices\cite{burch2018magnetism,doi:10.1021/acsnano.1c09150,zhong2017van,hu2023spin}. At a basic level, this coupling describes the dependence of magnetic couplings to the positions of atoms in the crystal, but may also include dynamical effects that couple spins to atomic momenta\cite{van1940paramagnetic,orbach1961spin,ray1967dynamical,capellmann1991spin,ioselevich1995strongly}.

Of recent fundamental interest are a number experimental findings, which all call for a deeper understanding of spin-lattice coupling. Example phenomena include: (i) The formation of magnon-polarons (hybrid one-magnon, one-phonon excitations), such as observed in (Y/Lu)MnO$_3$,\cite{petit2007spin,oh2016spontaneous} FePS$_3$,\cite{liu2021direct,vaclavkova2021magnon,zhang2021coherent,cui2023chirality} and Fe$_2$Mo$_3$O$_8$\cite{bao2023direct}. In the latter materials, the avoided crossings may give rise to topologically nontrivial magnetoelastic excitations\cite{to2023giant,luo2023evidence,bao2023direct}. (ii) Strong temperature dependence of the longitudinal (phonon) thermal conductivity and/or ultrasound attenuation\cite{zhou2011spinon,poirier2014ultrasonic,streib2015elastic,metavitsiadis2020phonon,feng2022sound} due to spin-phonon scattering, as observed in e.g.~YMnO$_3$\cite{kim2024thermal}, FeCr$_2$S$_4$\cite{zhou2024large}, $\alpha$-RuCl$_3$\cite{leahy2017anomalous, hentrich2018unusual}, and various MPS$_3$ (M = Fe, Mn) compounds\cite{haglund2019thermal}. (iii) The controversial phonon thermal Hall effect, now possibly observed in a wide variety of magnetic insulators including FeCr$_2$S$_4$\cite{zhou2024large}, FeCl$_2$\cite{xu2023thermal}, Fe$_2$Mo$_3$O$_8$\cite{ideue2017giant}, VI$_3$\cite{zhang2021anomalous}, CrI$_3$\cite{xu2024thermal}, YMnO$_3$\cite{kim2024thermal}, Na$_2$Co$_2$TeO$_6$\cite{gillig2023phononic,li2023magnon,chen2023planar,guang2023thermal} and $\alpha$-RuCl$_3$\cite{kasahara2018unusual, yokoi2021half, bruin2022robustness, czajka2023planar,lefranccois2022evidence}. It should be emphasized that the precise explanation of the thermal Hall effect in each material remains somewhat controversial, as there are a variety of potential contributions. These include extrinsic contributions (impurity skew-scattering)\cite{mori2014origin,guo2021extrinsic,guo2022resonant,li2023magnons} and intrinsic contributions from topological magnons\cite{onose2010observation,mcclarty2022topological,zhang2024thermal} and fractional spinons\cite{katsura2010theory,nasu2017thermal} in addition to chiral phonons. The latter may acquire chirality as a consequence of spin-lattice coupling, through topological magnon-polaron formation\cite{thingstad2019chiral,park2019topological,sheikhi2021hybrid,huang2021topological}, dissipationless spin-phonon scattering\cite{sheng2006theory,kagan2008anomalous,wang2009phonon,zhang2010topological,zhang2019thermal,saito2019berry,ye2020phonon,sun2021phonon,ye2021phonon,zhang2021phonon} (often captured by a phenomenological phonon Hall viscosity), and spin-phonon skew-scattering\cite{mangeolle2022phonon,mangeolle2022thermal}. In many cases, it has been argued that phonons are likely to contribute significantly to the observed thermal Hall conductivity\cite{zhang2021anomalous,lefranccois2022evidence,li2023magnons,gillig2023phononic,li2023magnon,chen2023planar,xu2023thermal}, but the precise mechanisms involved in each material remain unclear. What is currently missing is a comprehensive material-specific picture of spin-lattice coupling in each compound, and it's relationship to various properties.
Experimentally, it is clear that all of these compounds display strong spin-lattice coupling, in addition to having unquenched orbital moments and/or relatively anisotropic exchange interactions arising from spin-orbit coupling. Thus, we need a first principles approach capable of accurately treating the complex local spin-orbital moments and their coupling to phonons.  

In this work, we describe a new {\it ab-initio} method for addressing these questions in a material-specific way, by computing generic spin-phonon couplings within the framework of numerical (des Cloizeaux) effective Hamiltonians (dCEHs)\cite{des1960extension}. In the dCEH approach generally, a many-body electronic Hamiltonian is solved, and the resulting low-energy states are projected onto a reference low-energy space, in order to obtain an effective Hamiltonian that reproduces the low-energy spectrum exactly. The advantages of this approach are as follows: (i) The use of multi-reference wavefunctions allows the full many-body (spin-orbital multiplet) character of the local states to be faithfully captured, which is relevant for accurate calculations of magnetic anisotropies, and higher order interactions (e.g.~ring-exchange and biquadratic exchange), (ii) The method can be applied to any sufficiently local model, and can be easily adapted to treat any combination of spin, charge, and orbital degrees of freedom, (iii) The resulting couplings necessarily retain all symmetries of the paramagnetic crystal, and are quantitatively consistent with analytical perturbation theory by construction, (iv) The method is semi-empirical; a small number of parameters, such as (screened) Coulomb interaction strengths, serve as inputs. These parameters are taken from experimental estimates, when available. While the method is therefore not ``fully'' {\it ab-initio}, it does not require very large active spaces or many-body perturbation theory to properly capture the screening of the Coulomb interactions, which is a primary limitation of fully {\it ab-initio} multireference approaches. 

This paper is organized as follows. In Section \ref{sec:intuition} and \ref{sec:general}, we first motivate the form of the generic spin-lattice coupling from perturbation theory, and briefly review existing {\it ab-initio} approaches. In Section \ref{sec:dCEH}, we review des Cloizeaux effective Hamiltonians, and their implementation using cluster expansions. In Section \ref{sec:implementation}, we discuss the inclusion of phonons into this framework, and the extraction of spin-phonon couplings. 

We then apply this approach to study MnPSe$_3$. This material represents a simple case of strong spin-phonon coupling, as evidenced particularly by the Raman scattering experiment of Ref.~\onlinecite{mai2021magnon}, which shows strong renormalization of the optical phonon lifetimes due to hybridization with a two-magnon continuum. While this material does not display all of the spin-orbital complexity of the previously mentioned materials, it serves as an ideal benchmark for our new method. The more complex case of $\alpha$-RuCl$_3$ is addressed in a companion paper\cite{dhakal2024rucl3}. Thus, in Section \ref{sec:mn-intro} we provide an introduction to this material, and in Section \ref{sec:mn-hamiltonian} we compute the zeroth order magnetic couplings. Sections \ref{sec:phonons} and \ref{sec:mn-spin-phonon} address the phonons and spin-phonon couplings. Finally, in \ref{sec:mn-raman}, we combine the computed spin-phonon couplings with semiclassical spin dynamics to compute the magnetic and phononic Raman response for comparison with Ref.~\onlinecite{mai2021magnon}. We ultimately demonstrate quantitative agreement with the temperature-dependence of the phonon lifetimes. The results and perspective are summarized in Section \ref{sec:conclusions}.

\section{Spin-Phonon Couplings Theory}

\subsection{Intuition from Perturbation Theory}\label{sec:intuition}

In this section, we first motivate the form of the magnetoelastic couplings, with reference to perturbation theory in electron-phonon coupling. Seminal works on this topic\cite{van1940paramagnetic,orbach1961spin,ray1967dynamical,capellmann1991spin,ioselevich1995strongly} often only considered the effects of modulation of local crystal fields by phonons, which is likely the primary source of spin-phonon coupling in rare earth ions. For more general applications, it is instructive to consider modulation of both the crystal field and intersite hoppings. 
For the sake of discussion, we consider a two-orbital Hubbard model with spin-orbit coupling, with one electron per site on average. The base electronic Hamiltonian is given by:
\begin{align}
    \mathcal{H}_{\rm el} = & \ \mathcal{H}_0 + \mathcal{H}_{\rm hop} \\
    \mathcal{H}_0 = & \ \sum_{i\alpha} \epsilon_\alpha \mathbf{c}_{i\alpha}^\dagger \mathbf{c}_{i\alpha}  \nonumber \\ & \ + U\sum_i\left(n_{i,1}n_{i,2}+ \sum_{\alpha}n_{i,\alpha,\uparrow}n_{i,\alpha,\downarrow}\right)\\
    \mathcal{H}_{\rm hop} = & \  \sum_{ij\alpha\beta}\mathbf{c}_{i\alpha}^\dagger \mathbb{T}_{ij}^{\alpha\beta}\mathbf{c}_{j\beta}
\end{align}
where:
\begin{align}
    \mathbf{c}_{i,\alpha}^\dagger = \left(c_{i,\alpha,\uparrow}^\dagger \ \ c_{i,\alpha,\downarrow}^\dagger \right) \ \ \ , \ \ \ \mathbf{c}_{i,\alpha} = \left( \begin{array}{c}c_{i,\alpha,\uparrow} \\ c_{i,\alpha,\downarrow} \end{array}\right)
\end{align}
and $c_{i,\alpha,\sigma}^\dagger$ creates an electron at site $i$, orbital $\alpha\in\{1,2\}$, and spin $\sigma$. The orbital energies are taken to satisfy $\epsilon_2 > \epsilon_1$. The number operators are:
\begin{align}
    n_{i,\alpha} = \mathbf{c}_{i\alpha}^\dagger \mathbf{c}_{i,\alpha} = n_{i,\alpha,\uparrow} + n_{i,\alpha,\downarrow}
\end{align}
The hopping matrices are complex as a consequence of SOC, satisfy $\mathbb{T}_{ij}^{\alpha\beta} = (\mathbb{T}_{ji}^{\beta\alpha})^\dagger$, and can be written:
\begin{align}
    \mathbb{T}_{ij}^{\alpha\beta} = t_{ij}^{\alpha\beta} \mathbb{I}_{2\times 2} + i \vec{\lambda}_{ij}^{\alpha\beta} \cdot \vec{\sigma}
\end{align}
\begin{figure}[t]
\includegraphics[width=0.7\linewidth]{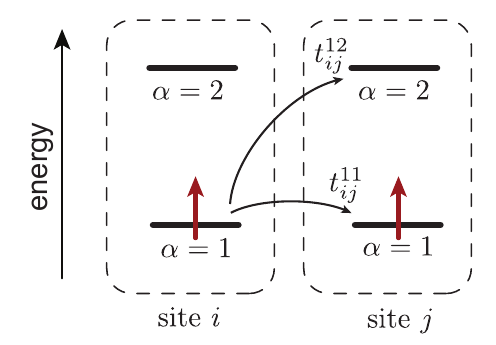}
\caption{Schematic definition of two-orbital model showing two sites, with orbital labels $\alpha,\beta\in\{1,2\}$, and hoppings $t_{ij}^{\alpha\beta}$ shown. The electronic ground state is taken to consist of one electron at each site occupying the $\alpha = 1$ orbital. }
\label{fig:schematic-twosite}
\end{figure}
where $\mathbb{I}_{2 \times 2}$ is a $2\times 2$ identity matrix. The linear electron-phonon coupling can be written:
\begin{align}
   \mathcal{H}_{\rm el-ph} =  & \ \sum_{i,\alpha\beta,n} \mathbf{c}_{i\alpha}^\dagger \mathbb{D}_{i,n}^{\alpha\beta}\mathbf{c}_{i,\beta} \ \hat{u}_n \nonumber \\ & \ + \sum_{ij,\alpha\beta,n} \mathbf{c}_{i\alpha}^\dagger \mathbb{M}_{ij,n}^{\alpha\beta}\mathbf{c}_{j,\beta} \ \hat{u}_n 
\end{align}
where $\hat{u}_n$ is a generic operator measuring the deviation of the atomic positions from equilibrium along some distortion vector. This may be written in terms of phonon operators as:
\begin{align}
    \hat{u}_n = & \ (a_n^\dagger + a_n) \\
    \hat{p}_n = & \ i\omega_n(a_n^\dagger - a_n) \\
    \mathcal{H}_{\rm ph} = & \ \frac{1}{4}\sum_n \hat{p}_n^2 + \omega_n\hat{u}_n^2 = \sum_n \omega_n\left(a_n^\dagger a_n + \frac{1}{2}\right)
\end{align}
Here, we have suppressed the units of the phonon operators for simplicity of notation; they will be restored in section \ref{sec:implementation} when discussing our {\it ab-initio} approach.
In $\mathcal{H}_{\rm el-ph}$, the first term $\mathbb{D}$ represents the modulation of the local crystal-field at site $i$, while the second term $\mathbb{M}$ represents the modulation of the intersite hopping. These may be written:
\begin{align}
    \mathbb{D}_{i,n}^{\alpha\beta} = & \ d_{i,n}^{\alpha\beta}\mathbb{I}_{2\times 2} + i\vec{\delta}_{i,n}^{\alpha\beta}\cdot \vec{\sigma}
    \\
    \mathbb{M}_{ij,n}^{\alpha\beta} = & \ m_{ij,n}^{\alpha\beta}\mathbb{I}_{2\times 2} + i\vec{\xi}_{ij,n}^{\alpha\beta} \cdot \vec{\sigma}
\end{align}
This explicitly distinguishes the spin-independent terms from the spin-dependent terms induced by SOC.

We consider the terms that arise perturbatively in intersite hopping $\mathcal{H}_{\rm hop}$ and electron-phonon coupling $\mathcal{H}_{\rm el-ph}$ (taking the unperturbed Hamiltonian to be $\mathcal{H}_0 + \mathcal{H}_{\rm ph}$). The low-energy manifold is taken to contain all states with exactly one electron per site occupying orbital $\alpha = 1$, as shown schematically in Fig.~\ref{fig:schematic-twosite}. We may then compute the effective Hamiltonian in this manifold using the standard Brillouin-Wigner approach:
\begin{align}\label{eqn:BW}
    \mathcal{H}_{\rm eff} = \mathbb{P}\mathcal{H}\mathbb{P} + \mathbb{P}\mathcal{H}\mathbb{Q}(\mathcal{E}-\mathbb{Q}\mathcal{H}\mathbb{Q})^{-1}\mathbb{Q}\mathcal{H}\mathbb{P} + ...
\end{align}
where $\mathcal{E}$ is the state energy, $\mathbb{P}$ is a projection operator onto the low-energy subspace, and $\mathbb{Q}=1-\mathbb{P}$.

Let us first consider the effects of modulation of the intersite hopping via the phonons. For this purpose, it is useful to note (for generic $\mathbb{A}_{ij} = a_{ij} + i\vec{\alpha}_{ij}\cdot\vec{\sigma}$ and $\mathbb{B}_{ij} = b_{ij} + i\vec{\beta}_{ij}\cdot \vec{\sigma})$ that:
\begin{align}\label{eqn:substitute}
    \mathbf{c}_i^\dagger \mathbb{A}_{ij} \mathbf{c}_j \mathbf{c}_j^\dagger & \mathbb{B}_{ji} \mathbf{c}_i   +(i\leftrightarrow j) =
    \nonumber \\
    & -4(a_{ij}b_{ji}+ \vec{\alpha}_{ij}\cdot \vec{\beta}_{ji})\left(\mathbf{S}_i\cdot\mathbf{S}_j\right)
    \nonumber \\
    &-4(b_{ji}\vec{\alpha}_{ij}-a_{ij}\vec{\beta}_{ji})\cdot\left(\mathbf{S}_i \times \mathbf{S}_j\right)
   \nonumber \\
    &+4 \ \mathbf{S}_i\cdot\left(\vec{\alpha}_{ij}\otimes\vec{\beta}_{ji} +\vec{\beta}_{ji}\otimes \vec{\alpha}_{ij}\right)\cdot\mathbf{S}_j
    \nonumber \\ 
     & +2i (a_{ij} \vec{\beta}_{ji}+b_{ji}\vec{\alpha}_{ij})\cdot \left(\mathbf{S}_i-\mathbf{S}_j\right)
     \nonumber \\
     &- 2i (\vec{\alpha}_{ij} \times \vec{\beta}_{ji})\cdot \left(\mathbf{S}_i+\mathbf{S}_j\right)
\end{align}
where:
\begin{align}
    S_{i,\alpha}^\mu = \frac{1}{2} \mathbf{c}_{i,\alpha}^\dagger \sigma_\mu \mathbf{c}_{i,\alpha}
\end{align}
are the different components of the spin operators for orbital $\alpha$ at site $i$, with $\mu\in\{x,y,z\}$. 

At linear order in electron-phonon coupling, we have two types of contributions. The first involves the modulation of the hopping between the occupied $\alpha = 1$ orbitals on each site. The second involves hopping from an occupied $\alpha = 1$ orbital on one site to an unoccupied $\alpha = 2$ orbital on another site. Together, these are:
\begin{widetext}
    \begin{align}
    \mathcal{H}_{\rm eff} =& \  \sum_{ij,n} \mathbf{c}_{i,1}^\dagger \mathbb{T}_{ij}^{11}\mathbf{c}_{j,1}(\mathcal{E}-\mathcal{H})^{-1}\mathbf{c}_{j,1}^\dagger \mathbb{M}_{ji,n}^{11}\mathbf{c}_{i,1}\hat{u}_n
     + \hat{u}_n\mathbf{c}_{i,1}^\dagger \mathbb{M}_{ij,n}^{11}\mathbf{c}_{j,1}(\mathcal{E}-\mathcal{H})^{-1}\mathbf{c}_{j,1}^\dagger \mathbb{T}_{ji}^{11}\mathbf{c}_{i,1}
     \nonumber \\
     & \ \ \ \ +\sum_{ij,n} \mathbf{c}_{i,1}^\dagger \mathbb{T}_{ij}^{12}\mathbf{c}_{j,2}(\mathcal{E}-\mathcal{H})^{-1}\mathbf{c}_{j,2}^\dagger \mathbb{M}_{ji,n}^{21}\mathbf{c}_{i,1}\hat{u}_n
     + \hat{u}_n\mathbf{c}_{i,1}^\dagger \mathbb{M}_{ij,n}^{12}\mathbf{c}_{j,2}(\mathcal{E}-\mathcal{H})^{-1}\mathbf{c}_{j,2}^\dagger \mathbb{T}_{ji}^{21}\mathbf{c}_{i,1}
    + (i\leftrightarrow j)
\end{align}
The inclusion of $\mathcal{H}_{\rm ph}$ in the denominator $(\mathcal{E}-\mathcal{H})$ has important consequences for the terms, because the energy $\mathcal{E}$ may differ from the excited state energy by the number of phonon quanta. Accounting for this leads to equation  (\ref{eqn:heff1}). Substituting equation (\ref{eqn:substitute}) into equation (\ref{eqn:heff1}), taking $U\gg \omega$, and keeping only spin-dependent terms then gives equation (\ref{eqn:heff2}):
    \begin{align}
    \mathcal{H}_{\rm eff}  = & \  -\sum_{ij,n} \left( \frac{a^\dagger_n}{U-\omega_n}+\frac{a_n}{U+\omega_n}\right) \mathbf{c}_{i,1}^\dagger \mathbb{T}_{ij}^{11}\mathbf{c}_{j,1}\mathbf{c}_{j,1}^\dagger \mathbb{M}_{ji,n}^{11}\mathbf{c}_{i,1}
    + \left( \frac{a^\dagger_n}{U+\omega_n}+\frac{a_n}{U-\omega_n}\right) \mathbf{c}_{i,1}^\dagger \mathbb{M}_{ij,n}^{11}\mathbf{c}_{j,1}\mathbf{c}_{j,1}^\dagger \mathbb{T}_{ji}^{11}\mathbf{c}_{i,1}
    \nonumber \\
    & \ \ \ \ \ \ \  -\sum_{ij,n} \left( \frac{a^\dagger_n}{U-\omega_n}+\frac{a_n}{U+\omega_n}\right) \mathbf{c}_{i,1}^\dagger \mathbb{T}_{ij}^{12} \mathbb{M}_{ji,n}^{21}\mathbf{c}_{i,1}
    + \left( \frac{a^\dagger_n}{U+\omega_n}+\frac{a_n}{U-\omega_n}\right) \mathbf{c}_{i,1}^\dagger \mathbb{M}_{ij,n}^{12}\mathbb{T}_{ji}^{21}\mathbf{c}_{i,1}
    + (i\leftrightarrow j) \label{eqn:heff1}
    \\
 = & \  \sum_{ij,n} \frac{8\hat{u}_n}{U}\left[\left(t_{ij}^{11}m_{ji,n}^{11}+ \vec{\lambda}_{ij}^{11}\cdot \vec{\xi}_{ji,n}^{11}\right)\left(\mathbf{S}_i\cdot\mathbf{S}_j\right)
+\left(m_{ji,n}^{11}\vec{\lambda}_{ij}^{11}-t_{ij}^{11}\vec{\xi}_{ji,n}^{11}\right)\cdot\left(\mathbf{S}_i \times \mathbf{S}_j\right)-  \mathbf{S}_i\cdot\left(\vec{\lambda}_{ij}^{11}\otimes\vec{\xi}_{ji,n}^{11} +\vec{\xi}_{ji,n}^{11}\otimes \vec{\lambda}_{ij}^{11}\right)\cdot\mathbf{S}_j\right]
 \nonumber \\
& \ +\sum_{a\in\{1,2\}}\frac{4\hat{p}_n}{U^2} \left[  (\vec{\lambda}_{ij}^{1a} \times \vec{\xi}_{ji,n}^{a1}-t_{ij}^{1a} \vec{\xi}_{ji,n}^{a1}-m_{ji,n}^{a1}\vec{\lambda}_{ij}^{1a})\cdot \mathbf{S}_i+(\vec{\lambda}_{ij}^{a1} \times \vec{\xi}_{ji,n}^{1a}+t_{ij}^{a1} \vec{\xi}_{ji,n}^{1a}+m_{ji,n}^{1a}\vec{\lambda}_{ij}^{a1})\cdot\mathbf{S}_j \right]
\label{eqn:heff2}
\end{align}
\end{widetext}
The first three terms in equation (\ref{eqn:heff2}) give the static modification to the Heisenberg, Dzyalloshinskii-Moriya, and symmetric anisotropic exchange that is linearly proportional to the displacement $\hat{u}_n$. These arise from the fact that intersite exchange couplings depend on the atomic positions. The terms arising proportional to $\hat{p}_n$ deserve special mention. These reflect the fact that the specific spin-orbital composition of the ground-state doublets at each site may be altered by the displacement of the atoms. To understand the consequences, it is useful to consider applying a strong magnetic field in order to split the doublet into $\psi_{i,1,\uparrow}(x_n)$ and $\psi_{i,1,\downarrow}(x_n)$ states. Suppose that $\psi_{i,1,\uparrow}(x_n)$ is the ground state. If the atoms are displaced very slowly (i.e. adiabatically), then the system will tend to stay in this state. This corresponds to the limit $\omega_n \to 0$, in which case the terms with $\hat{p}_n$ vanish. However, if the atoms are rapidly displaced to a new coordinate $x_n^\prime$, as long as $\langle \psi_{i,1,\uparrow}(x_n)|\psi_{i,1,\downarrow}(x_n^\prime)\rangle \neq 0$ the displacement may induce a transition from $\uparrow$ to $\downarrow$, corresponding to the application of $S_i^-$. From this intuition, one can expect the appearance of single-site spin operators multiplying the {\it time derivative} of the displacement, i.e.~$\propto \hat{p}_n$. Similar conclusions were drawn in Refs~\onlinecite{van1940paramagnetic,orbach1961spin,ray1967dynamical,capellmann1991spin,ioselevich1995strongly}, although in the context of two-phonon processes discussed below. 

Looking now at the hopping modulation terms second order with respect to the electron-phonon coupling, we have:
\begin{widetext}
    \begin{align}
    \mathcal{H}_{\rm eff} =& \  \sum_{ij,nm} \hat{u}_n\mathbf{c}_{i,1}^\dagger \mathbb{M}_{ij,n}^{11}\mathbf{c}_{j,1}(\mathcal{E}-\mathcal{H})^{-1}\mathbf{c}_{j,1}^\dagger \mathbb{M}_{ji,m}^{11}\mathbf{c}_{i,1}\hat{u}_m
     + \hat{u}_n\mathbf{c}_{i,1}^\dagger \mathbb{M}_{ij,n}^{12}\mathbf{c}_{j,2}(\mathcal{E}-\mathcal{H})^{-1}\mathbf{c}_{j,2}^\dagger \mathbb{M}_{ji,n}^{21}\mathbf{c}_{i,1}\hat{u}_m
    + (i\leftrightarrow j)
    \\
=& \  -\sum_{ij,nm} \left(\frac{a_n^\dagger a_m^\dagger}{U-\omega_n+\omega_m} +\frac{a_n^\dagger a_m}{U-\omega_n-\omega_m} + \frac{a_n a_m^\dagger}{U+\omega_n+\omega_m} + \frac{a_na_m}{U+\omega_n-\omega_m}\right)\left[\mathbf{c}_{i,1}^\dagger \mathbb{M}_{ij,n}^{11}\mathbf{c}_{j,1}\mathbf{c}_{j,1}^\dagger \mathbb{M}_{ji,m}^{11}\mathbf{c}_{i,1}\right.
\nonumber \\
& \ \hspace{80mm}\left.
     +\mathbf{c}_{i,1}^\dagger \mathbb{M}_{ij,n}^{12}\mathbb{M}_{ji,m}^{21}\mathbf{c}_{i,1}\right]
    + (i\leftrightarrow j)\label{eqn:heff3}
\end{align}
Taking the limit $U \gg \omega$, and keeping only spin-phonon coupling terms gives:
    \begin{align}
    \mathcal{H}_{\rm eff} 
=& \  \sum_{ij,nm} \frac{4\hat{u}_n\hat{u}_m}{U}
\left[
(m_{ij,n}^{11}m_{ji,m}^{11}+ \vec{\xi}_{ij,n}^{11}\cdot \vec{\xi}_{ji,m}^{11})\left(\mathbf{S}_i\cdot\mathbf{S}_j\right)
\right.
\nonumber \\
& \ \hspace{30mm}
+(m_{ji,m}^{11}\vec{\xi}_{ij,n}^{11}-m_{ij,n}\vec{\xi}_{ji,m}^{11})\cdot\left(\mathbf{S}_i \times \mathbf{S}_j\right)
- \mathbf{S}_i\cdot\left(\vec{\xi}_{ij,n}^{11}\otimes\vec{\xi}_{ji,m}^{11} +\vec{\xi}_{ji,n}^{11}\otimes \vec{\xi}_{ij,n}^{11}\right)\cdot\mathbf{S}_j
\nonumber \\
& 
+\sum_{ij,n,m\neq n} 
\frac{2}{U^2}\left(\hat{p}_n\hat{u}_m-\hat{u}_n\hat{p}_m\right)\sum_{a\in\{1,2\}}\left[
(\vec{\xi}_{ij,n}^{1a} \times \vec{\xi}_{ji,m}^{a1} - m_{ij,n}^{1a} \vec{\xi}_{ji,m}^{a1}-m_{ji,m}^{a1}\vec{\xi}_{ij,n}^{1a})\cdot \mathbf{S}_i\right.
\nonumber \\
& \ \hspace{80mm}\left.
+(\vec{\xi}_{ij,n}^{a1} \times \vec{\xi}_{ji,m}^{1a}+m_{ij,n}^{a1} \vec{\xi}_{ji,m}^{1a}+m_{ji,m}^{1a}\vec{\xi}_{ij,n}^{a1})\cdot \mathbf{S}_j
\right]\label{eqn:heff4}
\end{align}
\end{widetext}
Here, the first sum gives the static modification of the Heisenberg, Dzyalloshinskii-Moriya and symmetric anisotropic exchange that is quadratic in atomic displacements. The second sum in equation (\ref{eqn:heff4}) arises via a similar mechanism to the spin-momentum coupling in equation (\ref{eqn:heff2}). The modification of the spin-orbital composition of the ground state doublets with atomic position introduces terms related to the time derivative of those positions. Due to the antisymmetric combination of $\hat{u}$ and $\hat{p}$, this term resembles the coupling of the phonon angular momentum to the spin. Indeed, this is the microscopic origin of the phenomenological ``Raman'' spin-phonon coupling $\propto (\vec{x}\times \vec{p})\cdot\mathbf{S}$, which has been implicated in models of the phonon thermal Hall effect\cite{sheng2006theory,kagan2008anomalous,wang2009phonon,zhang2010topological,saito2019berry,sun2021phonon,ye2021phonon,zhang2021phonon}. However, as implied by equation (\ref{eqn:heff4}), the actual form of this coupling in real materials is sensitive to the microscopic details of the phonons and spin-orbit effects, such that it may be strongly anisotropic and dependent on the phonon mode. 

Finally, let us consider the effects of modulating the crystal field. The lowest order contribution arises at second order with respect to $\mathcal{H}_{\rm el-ph}$, and takes the form:
\begin{align}
    \mathcal{H}_{\rm eff} = \sum_{i,n,m} \hat{u}_n\mathbf{c}_{i,1}^\dagger \mathbb{D}_{i,n}^{12}\mathbf{c}_{i,2}(\mathcal{E}-\mathcal{H})^{-1}\mathbf{c}_{i,2}^\dagger \mathbb{D}_{i,m}^{21}\mathbf{c}_{i,1}\hat{u}_m
\end{align}
Expanding in the limit $\Delta_{12}= \epsilon_2-\epsilon_1\gg \omega$, and keeping only spin-dependent terms gives:
\begin{align}
    \mathcal{H}_{\rm eff} = & \ \frac{2}{\Delta_{12}^2}\sum_{i,n,m\neq n}
    (\hat{p}_n\hat{u}_m - \hat{u}_n \hat{p}_m ) \cdot \nonumber \\
    & \ \cdot(  \vec{\delta}_{i,n}^{12}\times \vec{\delta}_{i,m}^{21} - d_{i,n}^{12}\vec{\delta}_{i,m}^{21}-\vec{\delta}_{i,n}^{12}d_{i,m}^{21})\cdot \mathbf{S}_{i}\label{eqn:heff5}
\end{align}
Here, again, we find a contribution to the ``Raman'' spin-phonon coupling. It may be noted that the energy denominator may satisfy $\Delta_{12} \ll U$ for many materials, such that modulation of the crystal field makes the dominant contribution to the ``Raman'' term. This may be particularly true for systems with unquenched orbital momentum and weak SOC (such as first-row transition metals), as there may be very low-lying crystal field levels that can contribute to equation (\ref{eqn:heff5}).

While this simple two-orbital model is not sufficiently elaborate to model most real materials, it provides several insights: (i) In the absence of SOC, one can expect the spin-phonon coupling to modulate only the intersite Heisenberg coupling, through static terms proportional to $u$, $u^2$, ... This can be anticipated on the basis of symmetry analysis; the phonon displacements can coupling only to time-reversal even products of spin operators. (ii) In the presence of SOC, not only are the full intersite anisotropic exchange couplings modulated via static displacements $u$, $u^2$, ..., but also additional terms with phonon momentum $p$ arise due to modulation of the ground state spin-orbital composition. By symmetry, these terms must couple to odd order spin operators. 

Given the potential for significant complexity in the spin-phonon couplings in real materials, it is desirable to have a general numerical method for evaluating arbitrary spin-phonon couplings. 

\subsection{{General Form of Spin-Phonon Operators and Inadequacy of Static Approaches}} \label{sec:general}

Following from the previous section, the spin-phonon coupling for general $S$ takes the form:
\begin{widetext}
\begin{align}
    \mathcal{H}= & \ \sum_{i,\alpha\ell n} \left[\left(\mathbf{D}_i^{\alpha\ell n}\cdot\vec{\mathcal{O}}_i \right) \hat{u}_{\alpha}(\ell n)+\left( \mathbf{G}_i^{\alpha \ell n} \cdot \vec{\mathcal{O}}_i\right) \hat{p}_\alpha(\ell n)\right]
+\sum_{ij,\alpha\ell n}\left[ \left( \vec{\mathcal{O}}_i \cdot \mathbb{A}_{ij}^{\alpha \ell n} \cdot \vec{\mathcal{O}}_j\right)\hat{u}_\alpha(\ell n)
+\left( \vec{\mathcal{O}}_i \cdot \mathbb{B}_{ij}^{\alpha \ell n} \cdot \vec{\mathcal{O}}_j\right)\hat{p}_\alpha(\ell n)\right]
    \nonumber \\
    & \ + \sum_{i,\alpha\ell n, \beta\mathscr{k}  m}\left[
    \left(\mathbf{R}_i^{\alpha\ell n;\beta\mathscr{k} m}\cdot\vec{\mathcal{O}}_i \right) \hat{u}_{\alpha}(\ell n)\hat{u}_{\beta}(\mathscr{k}m)
    +\left(\mathbf{L}_i^{\alpha\ell n;\beta\mathscr{k} m}\cdot\vec{\mathcal{O}}_i \right) \hat{u}_{\alpha}(\ell n)\hat{p}_{\beta}(\mathscr{k}m)
    \right]
    \nonumber \\
    & \ + \sum_{ij,\alpha\ell n, \beta\mathscr{k}  m}\left[
\left(\vec{\mathcal{O}}_i \cdot \mathbb{K}_{ij}^{\alpha \ell n;\beta \mathscr{k} m} \cdot \vec{\mathcal{O}}_j \right)\hat{u}_{\alpha}(\ell n) \hat{u}_{\beta}(\mathscr{k}m)
+ \left(\vec{\mathcal{O}}_i \cdot \mathbb{N}_{ij}^{\alpha \ell n;\beta \mathscr{k} m} \cdot \vec{\mathcal{O}}_j \right)\hat{u}_{\alpha}(\ell n) \hat{p}_{\beta}(\mathscr{k}m)
\right]
\end{align}
\end{widetext}
where $\hat{u}_{\alpha}(\ell n)$ refers to the displacement from equilibrium of the atom in unit cell $\ell$ with index $n$, in the $\alpha \in \{x,y,z\}$ direction. The vector $\vec{\mathcal{O}}_i$ contains all local multipole operators appropriate for the local degrees of freedom. The couplings with boldface font, such as $\mathbf{D}_i^{\alpha\ell n}$, are vectors; the couplings with blackboard bold font, such as $\mathbb{A}_{ij}^{\alpha\ell n}$ are matrices.  As in the previous section, time-reversal symmetry restricts many terms to vanish. For example, $\mathbf{D}$ represents the linear modulation of the single-ion anisotropy terms with atomic displacement, and is thus finite for even multipole operators, but vanishes for all odd multipoles. If we consider only spin dipole operators $\vec{\mathcal{O}}_i \approx (S_i^x, S_i^y, S_i^z)$, then the form simplifies to:
\begin{align}
    \mathcal{H} = & \ \sum_{i,\alpha\ell n}\left( \mathbf{G}_i^{\alpha\ell n} \cdot \mathbf{S}_i\right) \hat{p}_\alpha(\ell n)
    \nonumber \\
    & \ + \sum_{ij,\alpha\ell n}\left(\mathbf{S}_i \cdot \mathbb{A}_{ij}^{\alpha \ell n} \cdot \mathbf{S}_j\right)\hat{u}_\alpha(\ell n) 
    \nonumber \\
    & \ + 
    \sum_{\begin{subarray}{c}ij \\ \alpha\ell n, \beta\mathscr{k}  m\end{subarray}}
\left(\mathbf{S}_i \cdot \mathbb{K}_{ij}^{\alpha \ell n;\beta \mathscr{k} m} \cdot \mathbf{S}_j \right)\hat{u}_{\alpha}(\ell n) \hat{u}_{\beta}(\mathscr{k}m)
    \nonumber \\
    & \ + \sum_{\begin{subarray}{c}i\\ \alpha\ell n, \beta\mathscr{k}  m\end{subarray}}\left(\mathbf{L}_i^{\alpha\ell n;\beta\mathscr{k} m}\cdot\mathbf{S}_i \right) \hat{u}_{\alpha}(\ell n)\hat{p}_{\beta}(\mathscr{k}m)
\end{align}
The estimation of $\mathbb{A}$ and $\mathbb{K}$ has been previously demonstrated using {\it static} first principles approaches, i.e.~frozen phonon and/or frozen magnon methods. Various approaches have been demonstrated for this purpose, and have been widely employed with notable successes in modelling e.g.~spin relaxation in single molecule magnets\cite{lunghi2017intra,albino2019first,guo2018magnetic,lunghi2017role,escalera2017determining,albino2019first,reta2021ab,lunghi2022toward}, and spin-phonon couplings in 2D magnets\cite{zhang2019first,delugas2023magnon,cui2023chirality}. For example, $\mathbb{A}$ and $\mathbb{K}$ can be computed from the differences between nuclear force matrices in different static magnetic configurations\cite{fennie2006magnetically,shen2008first,kumar2012spin,paul2015spin}. Alternatively, the changes in magnetic couplings for small static atomic displacements may be computed, either from the total electronic energies of multiple different magnetic configurations for each displacement\cite{paul2023ab,delugas2023magnon}, or magnetic force theorem approaches\cite{hellsvik2019general,sadhukhan2022spin,mankovsky2022angular,lange2023calculating}. The latter approach can also provide higher multipole-lattice couplings \cite{pi2014calculation,pi2014anisotropic}. A combined magneto-structural force theorem approach has also been reported\cite{mankovsky2023spin}, and a similar framework has been demonstrated that can address $\mathbf{G}$ and $\mathbf{L}$ terms by relating them to nuclear Berry curvatures\cite{bonini2023frequency,ren2024adiabatic}. To our knowledge, this latter approach is the first {\it ab-initio} method capable of treating the full range of spin-phonon couplings, provided the effects are accurately captured in the Kohn-Sham (single-determinant) energies. However, in order to complement these approaches, it is desirable to have a many-body method that fully captures the local many-body nature of the single-ion states, and is able to treat $\mathbf{G}$ and $\mathbf{L}$, which are experimentally relevant, for example, for the phonon thermal Hall effect\cite{sheng2006theory,kagan2008anomalous,wang2009phonon,zhang2010topological,saito2019berry,sun2021phonon,ye2021phonon,zhang2021phonon}. This can be accomplished by treating the phonons as genuine dynamical variables. In the next sections, we describe such a method, based on the des Cloizeaux effective Hamiltonian (dCEH) approach.

\subsection{{Review of des Cloizeaux Effective Hamiltonians}}\label{sec:dCEH}

In general, low-energy models constitute (i) an effective Hamiltonian in a reduced Hilbert space that replicates the low-energy spectrum of the full Hamiltonian, and (ii) a mapping between the degrees of freedom of the low-energy space and the full Hilbert space. In the context of magnetic couplings, the full Hilbert space is spanned by the electronic degrees of freedom, while the low-energy magnetic Hamiltonian is written in terms of spin operators. When using perturbation theory (as in the previous sections), the mapping between spin and electronic variables is well-defined, and the magnetic couplings are directly obtained in terms of the hopping and Coulomb parameters of the electronic model. However, use of perturbation theory becomes increasingly fraught as the electronic model becomes more complicated, and a convenient separation of the Hamiltonian into a zeroth order and perturbing term cannot be found. In such cases, it is advantageous simply to diagonalize the electronic Hamiltonian exactly (at least on a small number of sites), and identify the ``exact'' low-energy states directly. However, this only produces the local spectrum and low-energy states in terms of electronic degrees of freedom; a mapping to spin-variables is still necessary in order to obtain a low-energy Hamiltonian. 

There are various schemes to construct this mapping. For example, a popular method of estimating magnetic couplings from {\it ab-initio} calculations is total energy mapping using density functional theory (DFT) \cite{szilva2023quantitative}. In this approach, DFT calculations are performed with magnetic moments at each site suitably constrained to realise various different static magnetic patterns. The resulting energies are interpreted as corresponding to classical spin wavefunctions (i.e.~single electronic Slater determinant states, as required by the Kohn-Sham construction). 
While this is very successful in simple cases, more complex spin-orbital moments can sometimes lead to ambiguity in the mapping, because their local multi-determinant structure may not be precisely captured in the Kohn-Sham wavefunction. As an alternative, one can use explicit multi-determinant wavefunction-based methods (sometimes referred to as ``Quantum Chemistry'' approaches), in which case the mapping can be made on symmetry grounds\cite{katukuri2014kitaev,katukuri2014mechanism,yadav2016kitaev,bhattacharyya2022antiferromagnetic,pizzochero2020magnetic}, or by projection onto idealized low-energy states\cite{winter2016challenges,riedl2019ab}. Here we describe the latter approach, suggested originally by des Cloizeaux\cite{des1960extension}, and implemented in a wide variety of contexts\cite{neese2020effective}. Notable contributions to the calculation of magnetic Hamiltonians in this context have been made by Malrieu  \cite{calzado2002analysisa,calzado2002analysisb,durand2007effective,calzado2009analysis,malrieu2014magnetic} and Maurice \cite{maurice2009universal,maurice2010rigorous,maurice2010magnetostructural,maurice2013theoretical,ruamps2014interplay,maurice2016zero}.

The essential idea of des Cloizeaux Effective Hamiltonians (dCEHs) is to construct the mapping by requiring the true low-energy states to have maximal overlap with an idealized low-energy basis. For the purpose of computing magnetic couplings, we apply this approach to the results of exactly diagonalizing the electronic Hamiltonian on small clusters of sites, which are ultimately combined via a linked cluster expansion\cite{oitmaa2006series}. {\it Ab-initio} dCEHs have recently been utilized by the authors to study of quantum magnets with strongly anisotropic spin interactions \cite{winter2016challenges,winter2022magnetic,yang2023magnetic}, multi-spin ring-exchange \cite{riedl2019critical,riedl2021spin}, and complicated admixture of different $J_{\rm eff}$ states\cite{dhakal2024hybrid}. The steps are as follows. 
\begin{enumerate}
    \item Define the electronic Hamiltonian using appropriate {\it ab-initio} methods and/or input from experiments. This may include estimation of the hopping and crystal field terms via Wannier fitting of DFT calculations, fitting of experimental spectra, etc.
    \item Exactly diagonalize the electronic Hamiltonian on a finite number of sites to obtain the low-energy eigenstates $|\phi_n\rangle$ and energies $E_n$ in terms of the electronic degrees of freedom. The low-energy Hamiltonian is therefore:
    \begin{align}
        \mathcal{H}_{\rm low} = \sum_n E_n |\phi_n\rangle \langle \phi_n|
    \end{align}
    \item Define an idealized low-energy space of spin-orbital states in terms of the electronic degrees of freedom, $|\psi_n\rangle$. The projection operator onto the low-energy space is therefore $\mathbb{P} = \sum_n |\psi_n\rangle\langle \psi_n|$. This is essentially the same as the projection operator appearing in equation (\ref{eqn:BW}). For example, it's role is to project out double occupancies to arrive at a pure spin or spin-orbital state. 
    \item Rotate/project the low energy Hamiltonian into the idealized basis. Formally, this is accomplished via:
    \begin{align}
        \mathcal{H}_{\rm eff} = \mathbb{P}\mathbf{S}^{-1/2}  \mathcal{H}_{\rm low} \mathbf{S}^{-1/2}\mathbb{P}
    \end{align}
    where $\mathbf{S}$ is the overlap matrix of the projected low-energy states, with matrix elements:
    \begin{align}\label{eqn:overlapmatrix}
        \left[\mathbf{S}\right]_{nm} = \langle \phi_n| \mathbb{P}|\phi_m\rangle 
    \end{align}
The meaning of this expression is as follows. An intermediate basis can be constructed from the projection of the electronic states, $|\phi_n^\prime\rangle \equiv \mathbb{P}|\phi_n\rangle$. However, these $|\phi_n^\prime\rangle$ are not orthonormal. In order to orthonormalize them, we apply the symmetric (L\"owden) approach, $|\phi_n^{\prime\prime}\rangle = \mathbf{S}^{-1/2}|\phi_n^\prime\rangle$, where $[\mathbf{S}]_{nm} = \langle \phi_n^\prime | \phi_m^\prime\rangle$. By noting that $\mathbb{P}^2 = \mathbb{P}$, one arrives at equation (\ref{eqn:overlapmatrix}). The symmetric orthogonalization procedure has the feature that it preserves all symmetries, and maximizes $\sum_n\langle \phi_n^{\prime\prime}|\phi_n^\prime\rangle$. Once $\mathcal{H}_{\rm eff}$ is obtained numerically, the different matrix elements can then be analyzed to write the couplings in conventional forms in terms of multipole operators. 

\end{enumerate}
It may be noted that $\mathbf{S}^{-1/2}\mathbb{P}$ defines a unitary transformation that preserves the eigenvalues of $\mathcal{H}_{\rm low}$, so the resulting effective Hamiltonian is guaranteed to preserve the low-energy spectrum of the electronic model on the chosen cluster of sites, while being explicitly written in terms of low-energy operators. 

With the above procedure defined, the remaining question is how to combine dCEHs from different clusters of sites? Here, the insight is that the dCEHs for different clusters include different subsets of terms in a hypothetical perturbation theory expansion. For example, a two-site cluster will include all terms up to $t^2/U$, plus a subset of additional higher order contributions (where $t$ is the intersite hopping). A three-site cluster will include all terms up to $t^3/U^2$ plus a different subset of higher order contributions. By considering which terms are included on each cluster, it can be seen that the couplings in the low-energy Hamiltonians satisfy a convergent linked cluster expansion provided the standard perturbation theory in $t/U$ converges. In this expansion, the {\it essential} contributions of each cluster are summed, which are defined as the value of the coupling measured on that cluster, minus the essential contributions of all subclusters. For the smallest cluster on which that coupling may be defined (i.e.~when there are no subclusters), the essential contribution is equal to the value of the coupling on that cluster. If the cluster expansion is carried out to sufficiently large clusters, then it is possible to mitigate any spurious symmetry breaking induced by truncation of the clusters to a small number of sites.

\begin{figure}[t]
\includegraphics[width=0.9\linewidth]{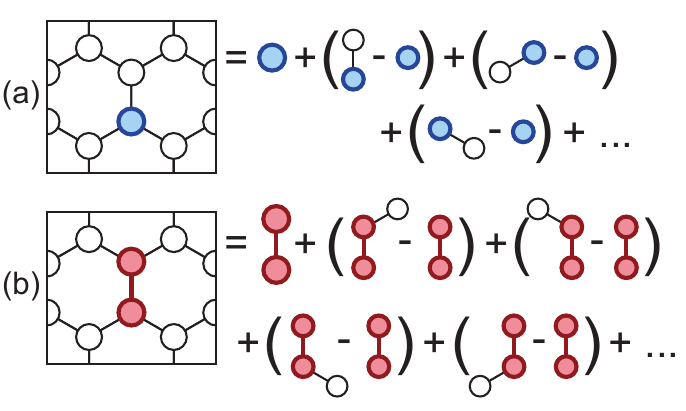}
\caption{Visualization of the lowest order terms in the cluster expansion for the nearest-neighbor honeycomb lattice for (a) one-site couplings and (b) two-site couplings. }
\label{fig:cluster-expansion}
\end{figure}

Fig.~\ref{fig:cluster-expansion} shows the pictorial representation of the lowest order contributions in the cluster expansions for one-site and two-site couplings. For example, consider a coupling defined on one site, such as a single-ion anisotropy term. The cluster expansion up to three-site clusters is:
\begin{align}
    D_i =& \  D_{i()} + \sum_j \left(D_{i(j)}-D_{i()}\right) \nonumber \\
    & \ + \sum_{jk}\left(D_{i(jk)}-D_{i(j)}-D_{i(k)}+D_{i()}\right) + ...
\end{align}
where the notation $D_{i(j)}$ denotes the value of $D_i$ obtained in the dCEH for a two site cluster with sites $i$ and $j$. Similarly, the cluster expansion for a bond interaction term can be written:
\begin{align}
    J_{ij} = & \ J_{ij()}+\sum_k \left( J_{ij(k)} - J_{ij()}\right) \nonumber \\
    & \ + \sum_{kl}\left(J_{ij(kl)}-J_{ij(k)}-J_{ij(l)}+J_{ij()}\right) + ...
\end{align}
In practice, however, it may not be necessary to consider clusters of more than two sites in each case. The point is that $J_{ij()} \propto t^2/U$, while the essential contribution from a three-site cluster is $(J_{ij(k)}-J_{ij()})\propto t^3/U^2$ at most. Provided the perturbation theory is well converging, the cluster expansion may converge rapidly with cluster size. It is always prudent to check larger clusters when computational resources allow, in order to confirm convergence. 

\subsection{{Spin-Phonon Implementation Details}}\label{sec:implementation}
In this section, we describe the inclusion of phonons explicitly in the dCEH approach.
We first compute the linear electron-phonon coupling in real space, defined by:
\begin{align}\label{eq:elphon}
\mathcal{H}_{\rm el-ph} = & \ \sum_{\alpha \ell n} \mathcal{H}_{\Delta}^{\alpha \ell n} \hat{u}_\alpha (\ell n)
\\ 
\mathcal{H}_{\rm \Delta}^{\alpha \ell n} =& \  \sum_{i\sigma j\sigma^\prime}\Delta_{i\sigma j\sigma^\prime}^{\mu\nu;\ell\alpha n}\  c_{i,\mu,\sigma}^\dagger c_{j,\nu,\sigma^\prime}
\end{align}
where $c_{i,\mu,\sigma}^\dagger$ creates an electron at metal site $i$, in orbital $\mu$, with spin $\sigma$. The phonon operator $\hat{u}_\alpha (\ell n)$ refers to displacement of atom $n$, in the unit cell labelled $\ell$, in the direction $\alpha \in \{x,y,z\}$. The elements $\Delta_{i\sigma j\sigma^\prime}^{\mu\nu;\ell\alpha n}$ refer to changes in the single-particle matrix elements due to such a displacement. In the present work, these are estimated via a finite displacement (frozen phonon) approach, as described in Section \ref{sec:electron-phonon}. Hopping matrix elements are computed using density functional theory for different atomic geometries, and $\Delta$ is determined from their differences.
Alternately, if one is interested in a small selection of phonon bands at specific $q$-points, the displacements may be taken along the relevant phonon eigenvectors.

We are now ready to compute the spin-phonon couplings. 
We first describe the calculation of $\mathbf{D}$, $\mathbf{G}$, $\mathbb{A}$, and $\mathbb{B}$,  which are linear in phonon operators. To include the effects of the real space displacements in the dCEH calculation of the magnetic couplings, we treat $u_{\alpha}(\ell j)$ and $p_\alpha(\ell j)$ formally in terms of fictitious phonon operators, given by:
\begin{align}
    a = & \ \sqrt{\frac{m_n \omega}{2\hbar}}\left(\hat{u}_{\alpha}(\ell n) + \frac{i}{m_n\omega} \hat{p}_\alpha(\ell n) \right)
    \\
    a^\dagger = & \ \sqrt{\frac{m_n\omega}{2\hbar}}\left(\hat{u}_{\alpha}(\ell n) - \frac{i}{m_n\omega} \hat{p}_\alpha(\ell n) \right)
    \\
    \hat{u}_{\alpha}(\ell n)=& \ \sqrt{\frac{\hbar}{2m_n\omega}}\left(a^\dagger + a\right) 
    \\
    \hat{p}_\alpha(\ell n) = & \ i \sqrt{\frac{\hbar m_n \omega}{2}} \left(a^\dagger - a\right)
\end{align}
In all calculations, we take:
\begin{align}
    r_0 \equiv & \ \sqrt{\frac{\hbar}{2m\omega}} = 0.02\text{ \AA}
    \\
    \hbar\omega = & \ 10\text{ meV}
\end{align}
As discussed below, provided electron-phonon coupling is in the perturbative regime, the computed spin-phonon couplings are linear in electron-phonon coupling. As a consequence, the specific values of the fictitious phonon constants are not important computationally, provided $\hbar \omega \ll U$ and $r_0 \Delta \ll U$, where $\Delta$ is the electron-phonon coupling defined in eq'n (\ref{eq:elphon}), and $U$ is the energy scale of the electronic excited states contributing at lowest order. The ficticious phonons are treated as hard-core bosons. A separate dCEH calculation is performed for each cluster included in the cluster expansion, and each atomic displacement defined by $\alpha,\ell,n$. The full Hamiltonian for each such calculation is constructed in the following way:
\begin{align}
    \mathcal{H} = \left(1 \ \ \ a^\dagger \right) \left(\begin{array}{c|c} 
    \mathcal{H}_{0} &  r_0 \ \mathcal{H}_\Delta^{\alpha \ell n} \\ \hline
    r_0 \ \mathcal{H}_\Delta^{\alpha \ell n} & \mathcal{H}_{0} + \hbar\omega \end{array} \right) \left(\begin{array}{c} \mathbb{P}_{n=0} \\ a \ \mathbb{P}_{n=1} \end{array} \right)
\end{align}
where $\mathcal{H}_{0}$ is the regular full electronic Hamiltonian including single-particle and Coulomb terms relevant for the cluster. 
 The Hilbert space is now doubled in size, and includes explicitly states with zero and one fictitious phonon. The projection operators $\mathbb{P}_{n=0}$ and $\mathbb{P}_{n=1}$ project onto subspaces with zero and one phonon, respectively. The idealized low-energy basis is also doubled in size; the projection operator can be constructed as:
\begin{align}
    \mathbb{P} = \left(\begin{array}{c|c} 
    \mathbb{P}_{\rm spin} & 0 \\ \hline
    0 & \mathbb{P}_{\rm spin}  \end{array} \right)
\end{align}
where $\mathbb{P}_{\rm spin}$ refers to the ideal low-energy space of the spin-only basis. The different blocks corresponds to the projection onto the sectors with nominally zero or one phonon.
Then, the low-energy Hamiltonian for each cluster can be numerically computed following section \ref{sec:dCEH}:
\begin{align}
    \mathcal{H}_{\rm eff} = & \  \mathbb{P}\mathbf{S}^{-1/2}\mathcal{H}_{\rm low}\mathbf{S}^{-1/2}\mathbb{P}
    \\
    = & \ \left(1 \ \ \ a^\dagger \right) \left(\begin{array}{c|c} 
    \mathcal{H}_{\rm eff}^{0} &     \mathcal{O}_{\alpha \ell n}  \\ \hline
    \mathcal{O}_{\alpha \ell n}^\dagger &   \mathcal{H}_{\rm eff}^{0}+  \mathcal{H}_{\rm eff}^{1} \end{array} \right)\left(\begin{array}{c} \mathbb{P}_{n=0} \\ a \ \mathbb{P}_{n=1} \end{array} \right)
\\
= & \ \mathcal{H}_{\rm eff}^0 + \mathcal{H}_{\rm eff}^1 \ a^\dagger a + \mathcal{O}_{\alpha\ell n}^\dagger \ a^\dagger + \mathcal{O}_{\alpha\ell n} \ a
\end{align}
where $\mathcal{H}_{\rm eff}^0$, $\mathcal{H}_{\rm eff}^1$, and $\mathcal{O}_{\alpha\ell n}$ act only on the spin degrees of freedom. 
For one-site clusters, 
\begin{align}
     \frac{\mathcal{O}_{\alpha \ell n}^\dagger+\mathcal{O}_{\alpha \ell n}}{2r_0} =& \  \mathbf{D}_{i()}^{\alpha\ell n}\cdot\vec{\mathcal{O}}_i  
     \\
     \frac{\mathcal{O}_{\alpha \ell n}^\dagger-\mathcal{O}_{\alpha \ell n}}{2i r_0 m_n \omega} = & \ \mathbf{G}_{i()}^{\alpha\ell n}\cdot\vec{\mathcal{O}}_i  
\end{align}
where $\mathbf{D}_{i()}^{\alpha \ell n}$ and $\mathbf{G}_{i()}^{\alpha \ell n}$ include all essential contributions in the cluster expansion from site $i$ (see section \ref{sec:dCEH} for cluster-expansion notation). For two-site clusters,
\begin{align}
         \frac{\mathcal{O}_{\alpha \ell n}^\dagger+\mathcal{O}_{\alpha \ell n}}{2r_0}  =& \  \mathbf{D}_{i(j)}^{\alpha\ell n}\cdot\vec{\mathcal{O}}_i  + \mathbf{D}_{j(i)}^{\alpha\ell n}\cdot\vec{\mathcal{O}}_j
         \nonumber \\
         & \ +  \vec{\mathcal{O}}_i \cdot \mathbb{A}_{ij()}^{\alpha \ell n} \cdot \vec{\mathcal{O}}_j
         \\
          \frac{\mathcal{O}_{\alpha \ell n}^\dagger-\mathcal{O}_{\alpha \ell n}}{2i \omega r_0 } = & \ m_n\left[\mathbf{G}_{i(j)}^{\alpha\ell n}\cdot\vec{\mathcal{O}}_i  + \mathbf{G}_{j(i)}^{\alpha\ell n}\cdot\vec{\mathcal{O}}_j\right.
          \nonumber \\
         & \ +  \left.\vec{\mathcal{O}}_i \cdot \mathbb{B}_{ij()}^{\alpha \ell n} \cdot \vec{\mathcal{O}}_j\right]
\end{align}
Then, up to two-site clusters, we have:
\begin{align}
    \mathbf{D}_i^{\alpha \ell n} \approx & \ \mathbf{D}_{i()}^{\alpha \ell n} + \sum_j \left(\mathbf{D}_{i(j)}^{\alpha \ell n} - \mathbf{D}_{i()}^{\alpha \ell n}\right)
    \\
    \mathbb{A}_{ij}^{\alpha \ell n} \approx & \ \mathbb{A}_{ij()}^{\alpha \ell n}
\end{align}
and analogous expressions for $m_n\mathbf{G}$ and $m_n\mathbb{B}$. At lowest order, all four sets of couplings are independent of the choice of $\hbar\omega$ and $r_0$. This can be seen from the fact that $(\mathcal{O}_{\alpha \ell n}^\dagger-\mathcal{O}_{\alpha \ell n}) \propto i\omega r_0$, and $(\mathcal{O}_{\alpha \ell n}^\dagger+\mathcal{O}_{\alpha \ell n}) \propto r_0$ at lowest order. For this reason, the specific choices of $\hbar\omega$ and $r_0$ do not affect the computed couplings. 

The computed spin-phonon couplings may then be Fourier transformed using phonon eigenvectors and frequencies derived from {\it ab-initio} phonon calculations. The displacement and momentum operators are written:
\begin{align}\label{eqn:uln}
\hat{u}_\alpha (\ell n) = \sqrt{\frac{\hbar}{2Nm_n}} \sum_{qv} \frac{e^{iq\cdot r_{\ell n}}}{\sqrt{\omega_{qv}} }\ ( a_{-qv}^\dagger+a_{qv})  \ e_\alpha(n,qv)\\ \label{eqn:pln}
\hat{p}_\alpha (\ell n) =i \sqrt{\frac{\hbar m_n}{2N} }\sum_{qv} \sqrt{\omega_{qv} }\ e^{iq\cdot r_{\ell n}} (a_{-qv}^\dagger - a_{qv} )  \ e_\alpha(n,qv)
\end{align}
where $q$ is the wavevector, $v$ is the phonon band index, $N$ = the number of unit cells, $m_n$ is the mass of atom $n$, and $e_\alpha$ are unitless coefficients appearing in the phonon eigenvectors. For example,
\begin{align}
    \sum_{i,\alpha\ell n} (\mathbf{D}_i^{\alpha \ell n} \cdot \vec{\mathcal{O}}_i) \hat{u}_\alpha (\ell n)
    \equiv \sum_{q\nu} \hat{\mathcal{D}}_{q\nu} (a_{-q\nu}^\dagger + a_{q\nu})
    \\
    \sum_{i,\alpha\ell n} (\mathbf{G}_i^{\alpha \ell n} \cdot \vec{\mathcal{O}}_i) \hat{p}_\alpha (\ell n)
    \equiv i\sum_{q\nu} \hat{\mathcal{G}}_{q\nu} (a_{-q\nu}^\dagger - a_{q\nu})
    \end{align}
where:
    \begin{align}
       \hat{\mathcal{D}}_{q\nu} = \sqrt{\frac{\hbar}{2N\omega_{q\nu}}}\sum_{i,\alpha\ell n}\sqrt{\frac{1}{m_n}} e^{iq\cdot r_{\ell n}}( \mathbf{D}_i^{\alpha \ell n} \cdot \vec{\mathcal{O}}_i) e_{\alpha}(n,q\nu)
       \\
   \hat{\mathcal{G}}_{q\nu} = \sqrt{\frac{\hbar\omega_{q\nu}}{2N}}\sum_{i,\alpha\ell n}\sqrt{\frac{1}{m_n}} e^{iq\cdot r_{\ell n}} (m_n \mathbf{G}_i^{\alpha \ell n} \cdot \vec{\mathcal{O}}_i)e_{\alpha}(n,q\nu)
\end{align}

To compute the $\mathbf{R}$, $\mathbf{L}$, $\mathbb{K}$, and $\mathbb{N}$ couplings, which are quadratic in phonon operators, we take a similar approach. For each cluster and pair of phonons $(\alpha\ell n) \neq (\beta\mathscr{k}m)$, we define:
\begin{align}
    \hat{u}_{\alpha}(\ell n)=& \ r_0\left(a_1^\dagger + a_1\right)
    \\
    \hat{p}_\alpha(\ell n) = & \ i \  r_0 \ m_1 \ \omega\left(a_1^\dagger - a_1\right)
    \\
        \hat{u}_{\beta}(\mathscr{k} m)=& \ r_0\left(a_2^\dagger + a_2\right)
    \\
    \hat{p}_\beta(\mathscr{k} m) = & \ i \  r_0 \ m_2 \ \omega\left(a_2^\dagger - a_2\right)
\end{align}
The Hamiltonian of the cluster is then:
\begin{align}
    \mathcal{H} =& \  \mathcal{H}_0 + \mathscr{a}^\dagger
    \left(\begin{array}{c|c|c|c} 
    0 & r_0\mathcal{H}_{\Delta,1} & r_0\mathcal{H}_{\Delta, 2} & 0 \\ \hline
    r_0\mathcal{H}_{\Delta,1} & \omega  & 0 & r_0 \mathcal{H}_{\Delta,2}\\ \hline
    r_0\mathcal{H}_{\Delta,2} & 0 &  \omega  & r_0\mathcal{H}_{\Delta,1} \\ \hline
    0& r_0\mathcal{H}_{\Delta,2} & r_0\mathcal{H}_{\Delta,1} &  2\omega 
    \end{array} \right) \mathscr{b}
\end{align}
where:
\begin{align}
    \mathscr{a}^\dagger = \left(1 \ \ a_1^\dagger \ \ a_2^\dagger \ \ a_1^\dagger a_2^\dagger\right) \ , \ 
    \mathscr{b} = \left(\begin{array}{c}
    \mathbb{P}_{n_1=0}^{n_2=0}\\
    a_1 \ \mathbb{P}_{n_1=1}^{n_2=0} \\
    a_2 \ \mathbb{P}_{n_1=0}^{n_2=1} \\
    a_1a_2 \ \mathbb{P}_{n_1=1}^{n_2=1}
    \end{array}\right)
\end{align}
Now, the Hilbert space explicitly includes states with zero or one quanta in each fictitious phonon mode. The low-energy projection operator is simply:
\begin{align}
    \mathbb{P} = \left(\begin{array}{c|c|c|c} 
    \mathbb{P}_{\rm spin} & 0 & 0 & 0\\ \hline
    0 & \mathbb{P}_{\rm spin} &0&0 \\ \hline
    0 & 0& \mathbb{P}_{\rm spin} &0 \\ \hline
    0 & 0 & 0 &\mathbb{P}_{\rm spin}
    \end{array} \right)
\end{align}
and the low-energy effective Hamiltonian is defined as:
\begin{align}
    \mathcal{H}_{\rm eff} = & \  \mathbb{P}\mathbf{S}^{-1/2}\mathcal{H}_{\rm low}\mathbf{S}^{-1/2}\mathbb{P}
    \\
    = & \ \mathcal{H}_{\rm eff}^0 + \mathcal{H}_{\rm eff}^1 n_1 + \mathcal{H}_{\rm eff}^2 n_2 + \mathcal{H}_{\rm eff}^3 n_1n_2 \nonumber \\ & + \mathscr{a}^\dagger\left(\begin{array}{c|c|c|c} 
   0 &     \mathcal{O}_{1} & \mathcal{O}_{2} & \mathcal{O}_{1;2}^{--}   \\ \hline
    \mathcal{O}_{1}^+ &    0 & \mathcal{O}_{1;2}^{+-} & \mathcal{O}_{1;2}^{0-}\\ \hline
    \mathcal{O}_2^+ & \mathcal{O}_{1;2}^{-+} & 0& \mathcal{O}_{1;2}^{-0} \\ \hline
    \mathcal{O}_{1;2}^{++} & \mathcal{O}_{1;2}^{0+} & \mathcal{O}_{1;2}^{+0} & 0
    \end{array} \right) \mathscr{b} 
    \label{eqn:expand1}
\end{align}
From this, we see for one-site clusters:
\begin{align}
    \frac{ \mathcal{O}_{1;2}^{++}+\mathcal{O}_{1;2}^{+-}+\mathcal{O}_{1;2}^{-+}+\mathcal{O}_{1;2}^{--}}{4r_0^2} =& \  \mathbf{R}_{i()}^{\alpha \ell n; \beta \mathscr{k} m}\cdot\vec{\mathcal{O}}_i
    \\
            \frac{ \mathcal{O}_{1;2}^{++}+\mathcal{O}_{1;2}^{+-}-\mathcal{O}_{1;2}^{-+}-\mathcal{O}_{1;2}^{--}}{4ir_0^2\omega} =& \  m_1\left[\mathbf{L}_{i()}^{ \beta \mathscr{k} m; \alpha \ell n}\cdot\vec{\mathcal{O}}_i\right]
            \\
        \frac{ \mathcal{O}_{1;2}^{++}-\mathcal{O}_{1;2}^{+-}+\mathcal{O}_{1;2}^{-+}-\mathcal{O}_{1;2}^{--}}{4ir_0^2\omega} =& \  m_2\left[\mathbf{L}_{i()}^{\alpha \ell n; \beta \mathscr{k} m}\cdot\vec{\mathcal{O}}_i\right]
\end{align}
and for two-site clusters:
\begin{align}
    \frac{ \mathcal{O}_{1;2}^{++}+\mathcal{O}_{1;2}^{+-}+\mathcal{O}_{1;2}^{-+}+\mathcal{O}_{1;2}^{--}}{4r_0^2} \hspace{30mm}
    \nonumber
    \\
      =\vec{\mathcal{O}}_i \cdot \mathbb{K}_{ij()}^{\alpha \ell n; \beta \mathscr{k} m}\cdot\vec{\mathcal{O}}_j + \mathbf{R}_{i(j)}^{\alpha \ell n; \beta \mathscr{k} m}\cdot\vec{\mathcal{O}}_i \hspace{10mm}
      \nonumber \\
      +\mathbf{R}_{j(i)}^{\alpha \ell n; \beta \mathscr{k} m}\cdot\vec{\mathcal{O}}_i
                  \\
          \frac{ \mathcal{O}_{1;2}^{++}+\mathcal{O}_{1;2}^{+-}-\mathcal{O}_{1;2}^{-+}-\mathcal{O}_{1;2}^{--}}{4ir_0^2\omega} \hspace{30mm}
    \nonumber
    \\
      =m_1\left[\vec{\mathcal{O}}_i \cdot \mathbb{N}_{ij()}^{ \beta \mathscr{k} m; \alpha \ell n}\cdot\vec{\mathcal{O}}_j + \mathbf{L}_{i(j)}^{\beta \mathscr{k} m; \alpha \ell n}\cdot\vec{\mathcal{O}}_i \hspace{10mm}\right.
      \nonumber \\
      \left.+\mathbf{L}_{j(i)}^{ \beta \mathscr{k} m; \alpha \ell n}\cdot\vec{\mathcal{O}}_i\right]
      \\
          \frac{ \mathcal{O}_{1;2}^{++}-\mathcal{O}_{1;2}^{+-}+\mathcal{O}_{1;2}^{-+}-\mathcal{O}_{1;2}^{--}}{4ir_0^2\omega } \hspace{30mm}
    \nonumber
    \\
      = m_2\left[\vec{\mathcal{O}}_i \cdot \mathbb{N}_{ij()}^{\alpha \ell n; \beta \mathscr{k} m}\cdot\vec{\mathcal{O}}_j + \mathbf{L}_{i(j)}^{\alpha \ell n; \beta \mathscr{k} m}\cdot\vec{\mathcal{O}}_i \hspace{10mm}\right.
      \nonumber \\
      \left.+\mathbf{L}_{j(i)}^{\alpha \ell n; \beta \mathscr{k} m}\cdot\vec{\mathcal{O}}_i\right]
\end{align}
Then, we have:
\begin{align}
    \mathbf{R}_i^{\alpha \ell n; \beta \mathscr{k} m} \approx & \ \mathbf{R}_{i()}^{\alpha \ell n; \beta \mathscr{k} m} + \sum_j \left(\mathbf{R}_{i(j)}^{\alpha \ell n; \beta \mathscr{k} m} - \mathbf{R}_{i()}^{\alpha \ell n; \beta \mathscr{k} m}\right)
    \\
    \mathbb{K}_{ij}^{\alpha \ell n; \beta \mathscr{k} m} \approx & \ \mathbb{K}_{ij()}^{\alpha \ell n; \beta \mathscr{k} m}
\end{align}
and analogous expressions for $\mathbf{L}$ and $\mathbb{N}$. It may be noted that the anharmonic terms involving three phonon $a^\dagger$ or $a$ operators in equation (\ref{eqn:expand1}) are expected to be small, and have thus been ignored in the later expressions.

Altogether, this method gives a recipe for the numerical estimation of generic spin-phonon couplings within an explicitly many-body framework, including both static and dynamic phonon terms. It can provide all multipole tensor interactions in the full $q$-space spin-phonon Hamiltonian, including modulation of single-ion anisotropies, $g$-tensors, intersite bilinear couplings, biquadratic couplings, etc. However, given the relatively large number of calculations in the cluster and phonon flavor expansion, it may appear to be very computationally heavy. In this context, it is worth noting that the most expensive step in dCEH calculations is the diagonalization of the combined electron phonon Hamiltonian on each cluster. It is typically possible to truncate the cluster expansion at two sites for magnetic insulators, and achieve quantitative results with small active spaces of orbitals for suitably constructed electronic Hamiltonians. For example, for transition metals, it is necessary to retain only the five $d$-orbital Wannier functions on each site; at half-filling and two phonon flavors, two-site clusters require diagonalization of a sparse matrix of dimension less than $10^6$, which can typically be accomplished within a few minutes on a single workstation without prohibitive memory requirements. The cluster expansion is easily parallelized. In practice, for moderately sized unit cells, the full spin-phonon dCEH calculation to achieve full $q$-dependent couplings  demands similar computational commitment as {\it ab-initio} phonon calculations. 

In order to validate the proposed method, in the next sections, we apply it to compute the linear spin-phonon couplings for MnPSe$_3$, and compare with experimental measurements of temperature- and band-dependent phonon renormalization.

\section{Study of $\text{MnPSe}_3$}

\subsection{Introduction} \label{sec:mn-intro}

\begin{figure}[t]
\includegraphics[width=0.9\linewidth]{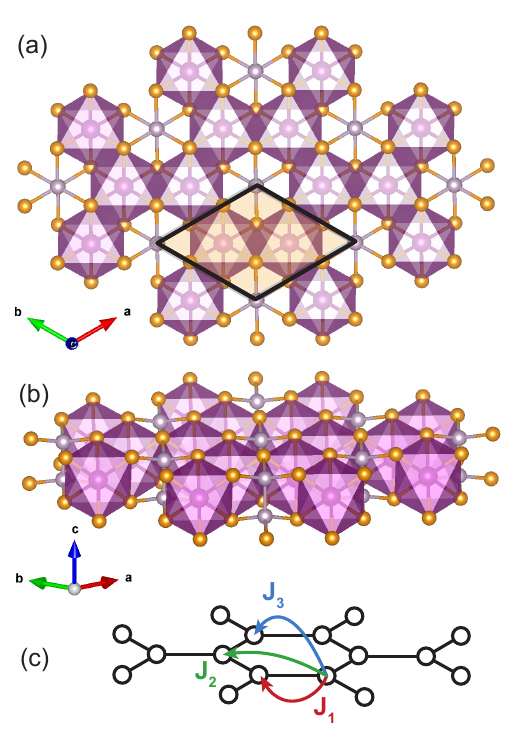}
\caption{Structure of MnPSe$_3$ monolayer viewed (a) along $c$-axis, with two-site honeycomb unit cell highlighted, and (b) from the side. (c) Definition of first, second, and third neighbor interactions between Mn sites. }
\label{fig:structure}
\end{figure}

MnPSe$_3$ is a layered van der Waals (vdW) material, with Mn sites forming a honeycomb lattice, as depicted in Fig.~\ref{fig:structure}. The material orders\cite{kim2019antiferromagnetic} as a N\'eel antiferromagnet below $T_N = 74$ K, below which there is a rapid increase in the longitudinal (phonon) thermal conductivity\cite{haglund2019thermal} and the lifetimes of some optical phonons\cite{mai2021magnon}, suggesting both are strongly influenced by spin-phonon scattering. Similar behavior is also observed in MnPS$_3$\cite{kim2019antiferromagnetic,vaclavkova2020magnetoelastic}. The detailed Raman scattering study of these effects in MnPSe$_3$ in Ref.~\onlinecite{mai2021magnon} provides the opportunity to benchmark our {\it ab-initio} approach via a comprehensive investigation of the magnetic couplings, lowest order spin-phonon couplings, and consequent temperature-dependent phonon lineshapes.

Given that the Mn$^{2+}$ in MnPSe$_3$ is in a high-spin $S = 5/2$ state, there is no significant orbital moment. As a consequence, the magnetic couplings are expected to be only weakly anisotropic, and spin-orbit coupling effects should be negligible. Powder neutron scattering\cite{calder2021magnetic} was shown to be compatible with the isotropic Heisenberg Hamiltonian containing couplings up to third neighbor:
\begin{align} \label{eqn:heisenbergham}
\mathcal{H} = \sum_{\langle ij \rangle } J_1  \ \mathbf{S}_i\cdot\mathbf{S}_j + \sum_{\langle \langle ij \rangle \rangle } J_2  \ \mathbf{S}_i\cdot\mathbf{S}_j
+ \sum_{\langle \langle \langle ij \rangle \rangle \rangle } J_3  \ \mathbf{S}_i\cdot\mathbf{S}_j
\end{align}
with values of $J_1 = 0.90$ meV, $J_2 = 0.06$ meV, and $J_3 = 0.38$ meV being motivated from previous {\it ab-initio} estimates\cite{sivadas2015magnetic,yang2020electronic} using DFT+$U$ total energy analysis.

It may be emphasized that this material, as a consequence of negligible SOC effects, should have a relatively simple form of the spin-phonon coupling. For this reason, it provides a ``simple'' material to benchmark our approach, even though it does not show the full range of couplings discussed in section \ref{sec:general}. In a subsequent paper, we address $\alpha$-RuCl$_3$, which displays more complex spin-orbital coupled moments\cite{dhakal2024rucl3}.

\subsection{Base Magnetic Hamiltonian} \label{sec:mn-hamiltonian}

\subsubsection{Treatment in Ab-initio} 
\label{sec:mnpse3_abinitio_treatment}

In order to develop a low-energy model for MnPSe$_3$, we first consider an electronic Hamiltonian in terms of Mn $3d$-orbitals, which is a sum of, respectively, one- and two-particle terms: $\mathcal{H}_{\rm el} = \mathcal{H}_{1p}+\mathcal{H}_{2p}$. The one-particle terms include intersite hopping, intrasite crystal field, and spin-orbit coupling, $\mathcal{H}_{1p} = \mathcal{H}_{hop}+\mathcal{H}_{\rm CF} + \mathcal{H}_{\rm SO}$:
\begin{align}
    \mathcal{H}_{hop} =& \  \sum_{ij\alpha\beta\sigma}t_{ij}^{\alpha\beta} c_{i,\alpha,\sigma}^\dagger c_{j,\beta,\sigma}
    \\
    \mathcal{H}_{\rm CF} = & \ \sum_{i\alpha\beta\sigma}d_{i}^{\alpha\beta}c_{i,\alpha,\sigma}^\dagger c_{i,\beta,\sigma}
    \\
    \mathcal{H}_{\rm SO} = & \ \sum_{i\alpha\beta\sigma\sigma^\prime} \lambda_{\rm Mn} \langle \phi_i^\alpha(\sigma)|\mathbf{L}\cdot\mathbf{S}|\phi_{i}^\beta(\sigma^\prime)\rangle c_{i,\alpha,\sigma}^\dagger c_{i,\beta,\sigma^\prime}
\end{align}
where $c_{i,\alpha,\sigma}^\dagger$ creates an electron at Mn site $i$, in $d$-orbital $\alpha$, with spin $\sigma$. For this purpose, we use $\lambda_{\rm Mn} = 41$ meV, which is the atomic value. As we see below, the high-spin $d^5$ configuration has no unquenched orbital momentum, such that SOC has only minor effects on the magnetic couplings. 
In order to estimate the hopping integrals ($\mathcal{H}_{hop}$) and crystalline electric field ($\mathcal{H}_{\rm CF}$), we started by taking a relaxed idealized monolayer structure with a vacuum gap of 11.98 \AA \ (see section \ref{sec:phonons} for details). The in-plane lattice parmater was found to be  6.345 \AA.  The non-spin polarized density functional theory (DFT) calculation was carried out at  Perdew–Burke–Ernzerhof generalized gradient approximation (PBE-GGA)\cite{perdew1996generalized} level as provided in a a full-potential local-orbital (FPLO) code\cite{koepernik1999full, opahle1999full}. We performed these calculations at non-relativistic limit by taking the dense k-point grid of $12\times 12\times12$. Finally, $\mathcal{H}_{hop}$ and ($\mathcal{H}_{\rm CF}$) terms were calculated by using Wannier projection, where the Wannier functions are assumed to have \textit{d}-only basis\cite{koepernik2023symmetry}.

For the two-particle terms, we consider both on-site and (long-range) intersite terms: $\mathcal{H}_{2p} = \mathcal{H}_U + \mathcal{H}_{J}^{\rm LR}$, respectively. The on-site contributions are given by:
\begin{align}
\mathcal{H}_U = \sum_{i\alpha\beta\delta\gamma}\sum_{\sigma\sigma^\prime}U_{\alpha\beta\gamma\delta} \ c_{i,\alpha,\sigma}^\dagger c_{i,\beta,\sigma^\prime}^\dagger c_{i,\gamma,\sigma^\prime} c_{i,\delta,\sigma}
\end{align}
In the spherically symmetric approximation \cite{sugano1970multiplets}, $U_{\alpha\beta\gamma\delta}$ are parameterized by the Slater parameters $F_0^{dd}, F_2^{dd}, F_4^{dd}$. For the (screened) on-site metal Coulomb interactions, we took $F_2^{dd} = 6.81$ eV, and $F_4^{dd} = 4.78$ eV, obtained from the fitting of low-temperature optical spectra of the related compound MnPS$_3$ in Ref.~\onlinecite{grasso1991optical}. Similar estimates were reported in other experimental studies\cite{grasso1989fluorescence,joy1992optical} for MnPS$_3$, and are not expected to differ significantly in MnPSe$_3$.

For the long-range intersite Hund's coupling, we define:
\begin{align}
    \mathcal{H}_{J}^{\rm LR} = \sum_{ij\alpha\beta\sigma\sigma^\prime} J_{H,ij}^{\alpha\beta} \ c_{i,\alpha,\sigma}^\dagger c_{j,\beta,\sigma^\prime}^\dagger c_{i,\alpha,\sigma^\prime}c_{j,\beta,\sigma}
\end{align}
The origin of this term is as follows. There are various exchange processes that contribute to the magnetic couplings, which have been discussed specifically in the context of MPS$_3$ materials\cite{sivadas2015magnetic,yang2020electronic,autieri2022limited}, as well as extensively for various other edge-sharing materials\cite{chaloupka2013zigzag,foyevtsova2013ab,liu2018pseudospin,liu2022exchange,liu2020kitaev,winter2022magnetic}.
If one considers explicitly the $d$-orbitals on the metal sites and the $p$-orbitals on the bridging ligands, the exchange processes can be grouped into two categories. The first category are those involving direct metal-metal hopping, giving $J \propto t_{dd}^2$. The second category are those involving metal-ligand hybridization, resulting in terms $J \propto t_{pd}^4$. This latter category includes the usual ferromagnetic Goodenough-Kanamori exchange\cite{goodenough1955theory,goodenough1958interpretation,kanamori1959superexchange}, which is typically important for edge-sharing compounds\cite{autieri2022limited}. All of these contributions can be downfolded, in principle, into the more convenient basis of metal-centered Wannier functions derived from the $d$-orbitals only. This reduced basis is more convenient from the standpoint of computational efficiency, but also because experimental on-site $F_2$ and $F_4$ parameters estimated from optical spectra refer to the Wannier basis implicitly.  The Wannier functions are, in reality, anti-bonding combinations of metal $d$- and ligand $p$-orbitals following standard ligand field theory. Since Wannier functions of different metal sites may share density on the same ligand, the downfolding produces longer-ranged intersite Coulomb terms. For the purpose of magnetic exchange, the long-range Hund's coupling should be considered explicitly -- this is the essence of the Goodenough-Kanamori mechanism. Here, we take a partially empirical approach to account for this contribution. Projecting the $p$-orbital Coulomb interactions into the $d$-orbital Wannier function basis gives the approximation:
\begin{align}
    J_{H,ij}^{\alpha\beta} = \gamma\sum_{n,\delta,\gamma}(U^n-J_H^n) & \ \phi_{i,\alpha}^{n,\delta}\phi_{j,\beta}^{n\delta}\phi_{i,\alpha}^{n,\gamma}\phi_{j,\beta}^{n,\gamma}\nonumber \\ & \ + J_H^n|\phi_{i,\alpha}^{n,\delta}|^2 |\phi_{j,\beta}^{n,\gamma}|^2 
\end{align}
where $\phi_{i,\alpha}^{n,\delta}$ is the wavefunction coefficient for the Wannier function at Mn site $i$ and $d$-orbital $\alpha$ corresponding to the Se atom $n$, and $p$-orbital $\delta \in\{p_x,p_y,p_z\}$. $U^n$ is the Hubbard repulsion between electrons in the same $p$-orbital at site $n$, and $J_H^n$ is the Hund's coupling at site $n$, and $\gamma$ is a renormalization factor that ostensibly accounts for screening of the interactions relative to their atomic values. We consider intersite Hund's coupling up to third neighbors, summing over all Se sites $n$ within 12\AA~of each of sites $i$ and $j$. We approximate $U^n = 2 J_H^n$, and take the selenium Hund's coupling to be the atomic value\cite{wittel1974atomic}, namely $J_H^n = 0.51$ eV. The wavefunction coefficients were taken from the DFT calculations of the hoppings, computed with FPLO as described above. A similar approach\cite{dhakal2024hybrid} was taken recently by the present authors in order to model the related compound FePS$_3$, where excellent agreement was demonstrated between computed and experimental excitations in the magnetically ordered phase. 

Finally, we vary the renormalization factor $\gamma$ and on-site repulsion $F_0^{dd}$ within a realistic range, and find good agreement with previous experimental couplings\cite{calder2021magnetic} for $\gamma = 0.65$, and $F_0^{dd} = 5.05$ eV. The latter value corresponds to $U_{t2g} = F_0^{dd} + (4/49)(F_2^{dd}+F_4^{dd}) = 6.0$ eV. The final  intersite Hund's matrices are given in Appendix \ref{sec:appendix}. 
At this point, it should be emphasized that this approach is not strictly fully {\it ab-initio}, since we have varied $U_{t2g}$ and $\gamma$ to best reproduce the experimental values of the base magnetic couplings in Ref.~\onlinecite{calder2021magnetic}. This is similar to the common practice of varying $U$ and/or $J_H$ in DFT+$U$ calculations of magnetic couplings to achieve reasonable experimental agreement. In the present case, $U_{t2g}$, and $\gamma$ control the relative contributions from antiferromagnetic and ferromagnetic exchange processes in addition to the overall scale of the couplings. Since we are primarily interested in studying the spin-phonon couplings, the utilization of experimental input in the selection of model parameters helps ensure accuracy of the resulting approach. These are the only two free parameters in our study, and all later magnetic and spin-phonon couplings derive from them. 

In order to determine the magnetic exchange constants, we exactly diagonalize the described electronic Hamiltonian on one-site and two-site clusters corresponding to first, second, and third neighbor bonds. For each cluster, we retain the lowest energy states, and then project the resulting low-energy Hamiltonian onto idealized pure-spin states representing the $^6A_1$ electronic ground states of each Mn site.

\subsubsection{Results}
Using the approach outlined in Section \ref{sec:mnpse3_abinitio_treatment}, we obtain the following estimates for the magnetic couplings. For a given bond, the bilinear couplings can be written:
\begin{align}
    \mathcal{H}=\sum_{ij}\mathbf{S}_i \cdot \mathbb{J}_{ij} \cdot \mathbf{S}_j
\end{align}
in terms of the $3\times 3$ exchange matrix $\mathbb{J}_{ij}$. For the nearest neighbor bond highlighted in Fig.~\ref{fig:structure}(c), we compute: 
\begin{align}
   \mathbb{J}_{1}=  \left(\begin{array}{ccc} 
0.9281 & 0.0002 & 0.0 \\ 
0.0002 & 0.9281 & 0.0 \\ 
0.0 & 0.0 & 0.9276 
 \end{array}\right)
\end{align}
which indeed shows very weak anisotropy. Similarly, for the second neighbor bond highlighted in Fig.~\ref{fig:structure}(c):
\begin{align}
   \mathbb{J}_2 =  \left(\begin{array}{ccc} 
0.0352 & -0.0007 & -0.0005 \\ 
0.0006 & 0.0352 & 0.0004 \\ 
0.0005 & -0.0004 & 0.0352 
 \end{array}\right)
\end{align}
Here, the nonzero off-diagonal components represent a nearly vanishing but symmetry-allowed Dzyalloshinkii-Moriya interaction, which we subsequently neglect due to it's exceedingly small magnitude. Finally, for the third neighbor bond highlighted in Fig.~\ref{fig:structure}(c), we estimate:
\begin{align}
    \mathbb{J}_3 = \left(\begin{array}{ccc} 
0.3791 & 0.0 & 0.0 \\ 
0.0 & 0.3791 & 0.0 \\ 0.0 & 0.0 & 0.3789 
 \end{array}\right)
\end{align}
From the same calculations, we also obtain higher order couplings, but can constrain all higher multipole couplings (such as biquadratic interactions) to be less than $10^{-5}$ meV. The single ion anisotropy terms are also found to be small, and take the form:
\begin{align}
    \mathcal{H}_{\rm SIA} =D  \sum_i   (S_i^c)^2
\end{align}
where we estimate $D = -2.6\times 10^{-4}$ meV, implying extremely weak Ising anisotropy. Such anisotropy is sufficiently weak that long-range dipolar interactions between the $S=5/2$ spins likely play a stronger role. Experimentally, MnPSe$_3$ shows weak easy-plane anisotropy. Neglecting these very small contributions, we find that the Hamiltonian does take the form of equation (\ref{eqn:heisenbergham}), with $J_1 = 0.93$ meV, $J_2 = 0.035$ meV, and $J_3 = 0.38$ meV. The computed couplings agree well with the values obtained from powder inelastic neutron scattering (partially by construction, as described above).

\begin{figure}[t]
\includegraphics[width=0.9\linewidth]{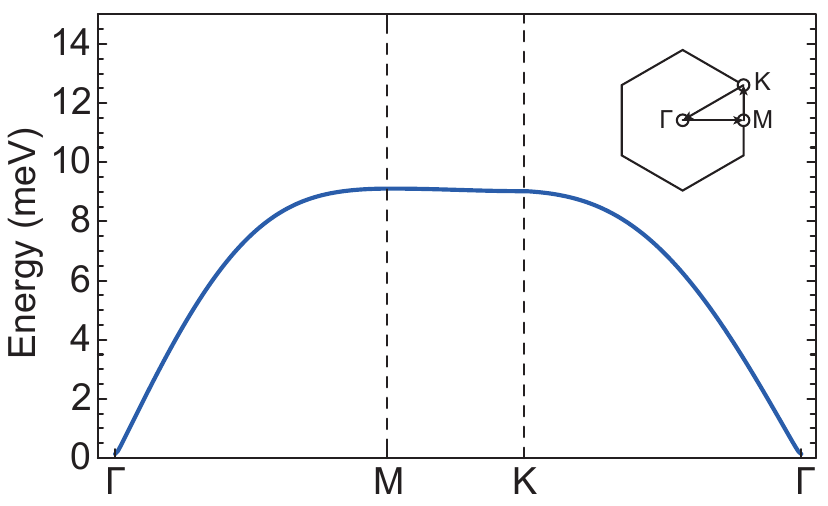}
\caption{Computed one-magnon dispersion for MnPSe$_3$ at the level of linear spin-wave theory. }
\label{fig:spinwave}
\end{figure}

In Fig.~\ref{fig:spinwave}, we show the expected magnon bands for these couplings at the level of linear spin-wave theory. Notably, the magnon bands are relatively dispersionless along the edge of the Brillouin zone at $\sim 9$ meV, which results in a large peak in the 2-magnon density of states (DOS) at $\sim 18$ meV. The latter peak is relevant to spin-phonon scattering, as discussed in subsequent sections. 

\subsection{Phonons}\label{sec:phonons}
To obtain the phonon eigenvectors, we carried out 
\textit{ab-initio} phonon calculations on a monolayer structure of MnPSe$_3$ using the Vienna \textit{Ab Initio} Simulation Package (VASP) \cite{kresse1996efficient,kresse1996efficiency,kresse1993ab} within the framework of DFT. The Projector augmented wave (PAW) pseudopotential  \cite{kresse1999ultrasoft,blochl1994projector} was chosen for all the elements as available in the VASP. The modified Perdew-Burke-Ernzerhof generalized gradient approximation functional  for solid (PBEsol GGA)\cite{perdew2008restoring} was used for the better approximation of exchange correlation potential. The DFT-D3 method of Grimme with zero-damping function\cite{grimme2010consistent} was implemented to account the correction due to van der Waals  forces. The kinetic energy cutoff for basis set and electronic self-consistent cycle (SCF) threshold was adjusted to 500 eV and 10$^{-8}$ eV respectively. A robust mixture of Blocked-Davidson  and RMM-DIIS iteration scheme \cite{wood1985new,pulay1980convergence} was used for electronic minimization algorithm. The spin polarized calculation was carried out with initial anti-ferromagnetic spin configuration in Mn sites and the electronic correlation was treated by implementing simplified rotationally invariant DFT+$U$ as suggested by Dudarev \textit{et al.}\cite{dudarev1998electron}. For this purpose, we applied a $U-J$ value of 4 eV  to the \textit{d}-orbitals of Mn atoms.  

\begin{figure}[t]
\includegraphics[width=0.9\linewidth]{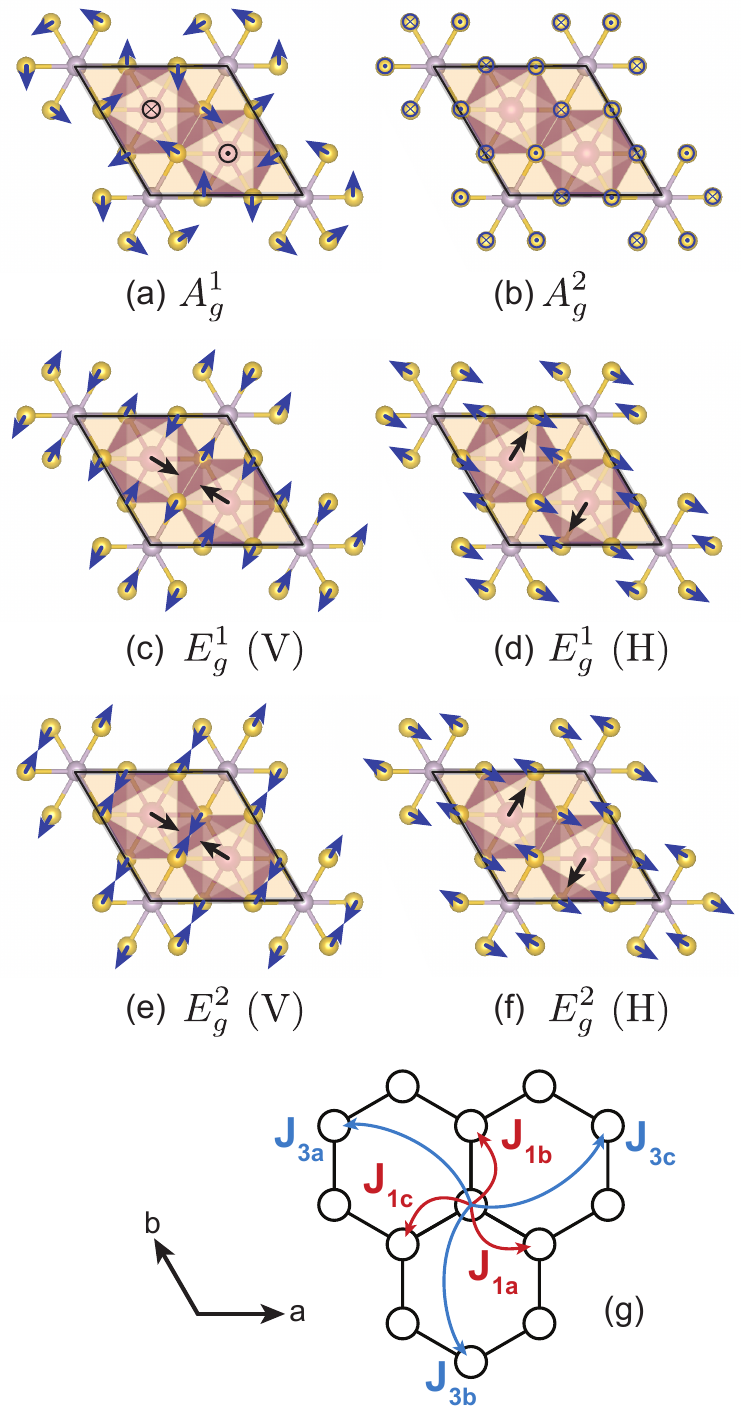}
\caption{(a-f) View of the displacements of the Mn and Se atoms for the lowest energy Raman-active optical $q=0$ phonons. For the $E_g$ modes, vertically (V) and horizontally (H) polarized combinations are indicated. (g) Definition of the individual first and third neighbor bonds referred to in Table \ref{tab:phonons}. }
\label{fig:phonons}
\end{figure}

To carry out phonon calculation, we started with structural relaxation of MnPSe$_3$ unit-cell without assuming any symmetry. The starting geometry was taken from the reported $R\bar{3}$ structure\cite{wiedenmann1981neutron}, with layers removed to produce a monolayer structure with a vacuum gap of $\sim 12$\AA \ between layers. The ionic relaxation was carried out until the forces became less than 0.001 eV/\AA. The Monkhorst-Pack grid of $12 \times 12 \times 4$ was taken for the structure relaxation purpose whereas less dense k-point grid of $6 \times 6 \times 4$ was chosen for the phonon calculation due to the supercell size of $2 \times 2 \times 1$. The phonon calculations were carried out by implementing finite displacement method\cite{kresse1995ab,parlinski1997first} with an atomic displacement of 0.02 \AA \ using Phonopy code\cite{phonopy-phono3py-JPCM,phonopy-phono3py-JPSJ}.   

In Table \ref{tab:phonons}, we show a comparison of the computed $q = 0$ monolayer phonon energies for the lowest Raman active modes with bulk experimental energies from Ref.~\onlinecite{mai2021magnon}. A reasonable agreement is found, despite taking a monolayer structure. In Fig.~\ref{fig:phonons}, we show the atomic displacements associated with such phonon modes. The $A_g^1$ phonon corresponds to out-of-plane modulation of the Mn atoms, and rotations of the P$_2$Se$_6$ anions. The $A_g^2$ phonon is a symmetric P$_2$Se$_6$ stretching mode that does not involve the Mn atoms. The $E_g^1$ and $E_g^2$ modes correspond to in-plane motions of the Mn atoms coupled to bending modes of the P$_2$Se$_6$ anions.

\begin{table*}[]
    \caption{\label{tab:phonons}Summary of lowest energy Raman active phonons and their computed $q=0$ spin-phonon couplings. For the doubly degenerate $E_g$ modes, the phonons have been symmetrized into horizontal (H) and vertical (V) polarizations. See Fig.~\ref{fig:phonons}(g) for definition of bonds.}
    \centering
    \begin{ruledtabular}
	\begin{tabular}{c|cc|ccccccc}
 Phonon & \multicolumn{2}{c|}{Phonon Energy (cm$^{-1}$)}& \multicolumn{7}{c}{Computed $q = 0$ Spin-Phonon Couplings (meV)}\\
		Label & Experiment\cite{mai2021magnon}& Computed (Monolayer) & Pol. &  $A_{1a}$ & $A_{1b}$ & $A_{1c}$  & $A_{3a}$ & $A_{3b}$ & $A_{3c}$                       \\  \hline
  		$A_g^1$          & 48.3      &    33.0 & & $\approx 0$& $\approx 0$& $\approx 0$& $\approx 0$& $\approx 0$& $\approx 0$     \\
		$E_g^1$          & 84.0    &     74.3 
  & V &+0.437 & -0.218 & -0.218 & +0.020 & -0.010 & -0.010   \\ 
  && & H & 0 & +0.378 & -0.378 & 0 & +0.017 & -0.017   \\
		$E_g^2$ & 109.6      &   103.1  
  & V &+0.502 & -0.251 & -0.251 & +0.121 & -0.061 & -0.061   \\ 
  && & H & 0 & +0.434 & -0.434 & 0 & +0.105 & -0.105   \\
		$A_g^2$          & 149.0     &    138.9 && $\approx 0$& $\approx 0$& $\approx 0$& $\approx 0$& $\approx 0$& $\approx 0$         \\ 
	\end{tabular}%
 \end{ruledtabular}
\end{table*}

\subsection{Spin-Phonon Couplings}\label{sec:mn-spin-phonon}
\subsubsection{Treatment in ab-initio}\label{sec:electron-phonon}

In order to model the Raman scattering experiment of Ref.~\onlinecite{mai2021magnon}, we focus on the $q=0$ phonons. For this purpose, we need only the electron-phonon coupling operator of equation (\ref{eq:elphon}) at $q=0$, which can be computed without use of supercells. Therefore, to estimate the $\Delta_{i\sigma j\sigma^\prime}^{\mu\nu;\ell\alpha n}$ parameters, we employed a finite difference (frozen phonon method). DFT calculations were performed with FPLO on a series of geometries with each atom displaced $\pm 0.02$\AA \ along each of the cartesian directions $\alpha \in \{x,y,z\}$. For each calculation, the hoppings were extracted via Wannier interpolation. The parameters of each calculation were otherwise identical to those described in Section \ref{sec:mnpse3_abinitio_treatment}. The linear electron-phonon couplings were then estimated by taking the differences between calculations displaced in the $+\alpha$ and $-\alpha$ directions. In the future, such calculations could employ more sophisticated approaches, such as those implemented in the EPW code\cite{giustino2007electron,noffsinger2010epw,lee2023electron}. 

In order to compute the linear spin-phonon couplings $\mathbb{A}$, $\mathbb{B}$, $\mathbf{D}$, and $\mathbf{G}$, we followed the approach outlined in section \ref{sec:implementation}. One- and two-Mn site clusters were exactly diagonalized, and the low-energy states were projected onto the $^6A_1$ electronic ground states. 

\subsubsection{$\Gamma$-point Results}

Here, we consider the spin-phonon couplings linear in displacement $u$ for $q=0$ phonons, which is relevant for modelling the Raman measurements in Ref.~\onlinecite{mai2021magnon}. We found numerically that the couplings $\mathbf{D}, \mathbf{G}$, and $\mathbb{B}$ are vanishing ($<10^{-5}$ meV/\AA) for MnPSe$_3$, as expected due to the weak effect of SOC. However, the couplings $\mathbb{A}$ are large. It is instructive to consider these couplings in mixed real-momentum space representation. The relevant coupling for the $q=0$ phonons is:
\begin{align}
    \mathcal{H}_{sp} = \sum_{(ij)}\sum_\nu \mathcal{A}_{\nu,{q=0}} \ (a_{\nu,q  =  0}^\dagger + a_{\nu,q=0})
\end{align}
The spin-phonon coupling operator is:
\begin{align}
    \mathcal{A}_{\nu,q=0} =& \  \frac{1}{\sqrt{N}}\sum_{\rm u.c.}^{N} \sum_{(ij)} \mathbf{S}_i \cdot \mathbb{A}_{(ij)}^{\nu,q=0}\cdot \mathbf{S}_j
\end{align}
where the summation over $(ij)$ refers to the six types of first and third neighbor bonds in Fig.~\ref{fig:phonons}(g). The summation over u.c. refers to unit cells. $\mathbb{A}_{(ij)}^{\nu,q=0}$ gives the change in the magnetic coupling between spins $i$ and $j$ associated with the motion of the atoms along the coordinate of phonon band $\nu$ and momentum $q=0$. The $q$-space couplings for a given bond are related to the real-space couplings via:
\begin{align}
    \mathbb{A}_{(ij)}^{\nu,q=0} = \sum_{\alpha \ell n} \sqrt{\frac{\hbar}{2m_n\omega_{qv}}} \  \mathbb{A}_{(ij)}^{\alpha\ell n} \ e_{\alpha}(n,q=0,\nu)
\end{align}
where $m_n$ is the mass of the $n$th nucleus, $\omega_{q\nu}$ is the phonon frequency, and $e_{\alpha}(n,q\nu)$ is the phonon eigenvector coefficients.

The results for different phonon bands are summarized in Table \ref{tab:phonons}. As might be expected from the zeroth order magnetic Hamiltonian, the spin-phonon coupling matrices $\mathbb{A}_{(ij)}^{\nu,q=0}$ take the form of Heisenberg interactions for all phonon bands. For this reason, we present only the diagonal element of $\mathbb{A}$ in Table \ref{tab:phonons}, denoted $A_{(ij)}^{v,q=0}$. There are a few immediate observations. 

\begin{enumerate}

    \item The motion of the atoms, in principle, modulates both the hoppings and the intersite Hund's coupling terms. However, in the present case, we find that the contribution to $A_{(ij)}^{\nu,q=0}$ from the intersite Coulomb terms is small, such that it can be neglected at first approximation. This is because the intersite Coulomb tensors are not strongly modified by small atomic displacements. Instead, it is the (antiferromagnetic) kinetic exchange contributions associated with direct metal-metal hopping that are most strongly modulated.

    \item In the high-spin $d^5$ configuration of Mn$^{2+}$, all $d$-orbitals are singly occupied, such that the spin-density is essentially spherically symmetric around the Mn atom. The exchange contributions associated with direct metal-metal hopping are therefore primarily sensitive to the absolute Mn--Mn distances, rather than the relative orientations of the MnSe$_6$ octahedra. As such, the relative magnitudes of particular $A_{(ij)}^{\nu,q=0}$ follow from the amount to which the phonon mode $\nu$ modulates a particular bond distance $|\vec{r}_i - \vec{r}_j|$. For example, the $A_g^1$ and $A_g^2$ modes are not associated with in-plane motion of the Mn atoms (see Fig.~\ref{fig:phonons}). These phonons do not significantly modulate any of the Mn-Mn bond distances at linear order, and therefore do not couple strongly to the spins. 

    \item The spin-phonon coupling associated with the third neighbor bonds is significantly weaker than might be anticipated from the zeroth order $J_1/J_3$ ratio. For third neighbors, the primary exchange pathway involves ligand-mediated hoppings exclusively, which are modulated less than the nearest neighbor direct metal-metal hoppings. As a consequence, the spin-phonon coupling is largely dominated by nearest neigbor contributions. This feature may also tend to enhance spin-phonon coupling in edge-sharing or face-sharing geometries, in comparison to edge-sharing geometries. For the latter case, direct metal-metal hopping is typically weak due to the larger distance between metal ions.
\end{enumerate}

\subsection{{Modelling Raman Scattering}}\label{sec:mn-raman}

In order to validate the computed spin-phonon couplings, we investigated theoretically the temperature-dependence of the $q=0$ phonon lineshapes in MnPSe$_3$, observed in recent Raman scattering experiments\cite{mai2021magnon}. The Raman spectra contain contributions from both phonons and two-magnon excitations via conventional Fleury-Loudon scattering\cite{fleury1968scattering}. The intensity of the latter may be approximated by:
\begin{align}
    I_{\rm 2mag}(\omega) \propto \int dt \ e^{i\omega t}\langle \hat{\mathcal{R}}(t) \hat{\mathcal{R}}(0)\rangle \label{eqn:bond-bond1}
\end{align}
where the Raman operators $\hat{\mathcal{R}}$ depend on the polarization of the incoming and outgoing light $\hat{e}_{\rm in}$ and $\hat{e}_{\rm out}$, according to:
\begin{align}
    \hat{\mathcal{R}} \propto \sum_{ij} J_{ij} \  \mathbf{S}_i \cdot\mathbf{S}_j \ (\vec{r}_{ij} \cdot \hat{e}_{\rm in})(\vec{r}_{ij} \cdot \hat{e}_{\rm out})
\end{align}
where $\vec{r}_{ij}$ is the vector between sites $i$ and $j$. In the absence of strong SOC effects, Raman scattering probes bond-bond correlations, which are primarily sensitive to two-magnon excitations in colinear antiferromagnets. Strictly speaking, given the different contributions to the intersite exchange, the precise proportionality of the bond operators in $\hat{\mathcal{R}}$ to the intersite $J_{ij}$ does not hold (see Ref.~\onlinecite{yang2021non}). However, for MnPSe$_3$, this is a sufficiently reasonable approximation.

\begin{figure*}[t]
\includegraphics[width=\linewidth]{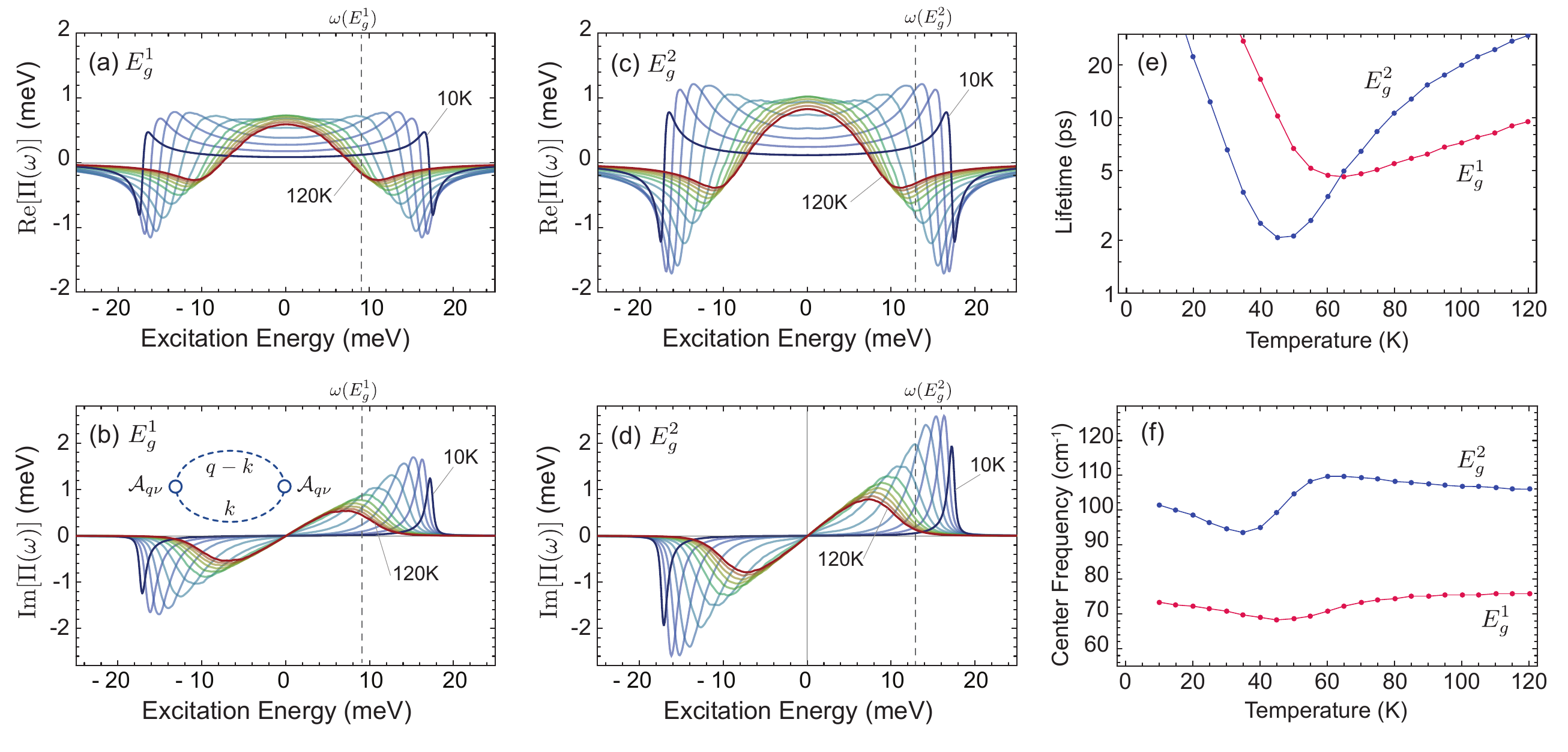}
\caption{(a-d) Computed $q=0$ phonon self-energy as a function of temperature: (a) Re[$\Pi(\omega+i\eta)$] for $E_g^1$, (b) Im[$\Pi(\omega+i\eta)$] for $E_g^1$, (c) Re[$\Pi(\omega+i\eta)$] for $E_g^2$, (d) Im[$\Pi(\omega+i\eta)$] for $E_g^2$. Curves are drawn for every 10 degrees from 10 K to 120 K. The bare phonon frequency of each mode, computed from {\it ab-initio}, is indicated. The inset of (b) shows the one-loop self-energy diagram associated with decay of a phonon with momentum $q$ into two-magnon excitations with momenta $q-k$ and $k$. (e) Computed phonon lifetime $\tau_\nu$. (f) computed center frequency $\tilde{\omega}_\nu$.}
\label{fig:selfenergy}
\end{figure*}

By virtue of the form of the first order spin-phonon coupling, the scattering of the phonons by spin excitations also generates a contribution to the phonon self-energy related to bond-bond spin correlations. The phonon Greens function is given by:
\begin{align}
    \mathscr{D}_{q,\nu}(i\nu_n) = \frac{1}{[\mathscr{D}_{q,\nu}^0(i\nu_n)]^{-1} - \Pi_{q,\nu}(i\nu_n)}
\end{align}
where $\nu_n = 2n\pi k_BT$, the phonon self-energy is $\Pi_{q,\nu}$ and the zeroth order Greens function is:
\begin{align}
   \mathscr{D}_{q,\nu}^0(i\nu_n) = \frac{1}{i\nu_n-\omega_{q\nu}^0}-\frac{1}{i\nu_n+\omega_{q\nu}^0}
\end{align}
where $\omega_{q\nu}^0$ is the bare phonon frequency. The phonon contribution to the Raman intensity is then:
\begin{align}\label{eq:phononlineshape}
    I_{ph}(\omega) \propto \sum_{\nu} [1+n_B(\omega)]|R_\nu|^2 B_{0, \nu}(\omega)
\end{align}
where $n_{B}(\omega) = (e^{\beta\omega}-1)^{-1}$ is the Bose distribution, $\beta = 1/(k_BT)$, and $|R_\nu|^2$ is the band-dependent Raman phonon intensity. $B_{0, \nu}(\omega)$ is the phonon spectral density, given by:
\begin{align}
    B_{q,\nu}(\omega) =-\frac{1}{\pi}\text{Im}\left[\mathscr{D}_{q,\nu}(i\nu_n)|_{i\nu_n\to \omega+i\Gamma} \right]
\end{align}
where $\Gamma$ is a bare broadening (unrelated to spin-phonon scattering). 
Accounting for the lowest order spin-phonon scattering diagram shown in the inset of Fig.~\ref{fig:selfenergy}(b), the band-dependent phonon self-energy is:
\begin{align}\label{eqn:bond-bond2}
    \Pi_{q,\nu}(i\nu_n) = -\int_0^\beta d\tau \ e^{i\nu_n \tau} \langle \mathcal{A}_{q\nu}(\tau)\mathcal{A}_{-q\nu}(0)\rangle
\end{align}
This represents the decay of phonons into two-magnon excitations.

\begin{figure}[t]
\includegraphics[width=\linewidth]{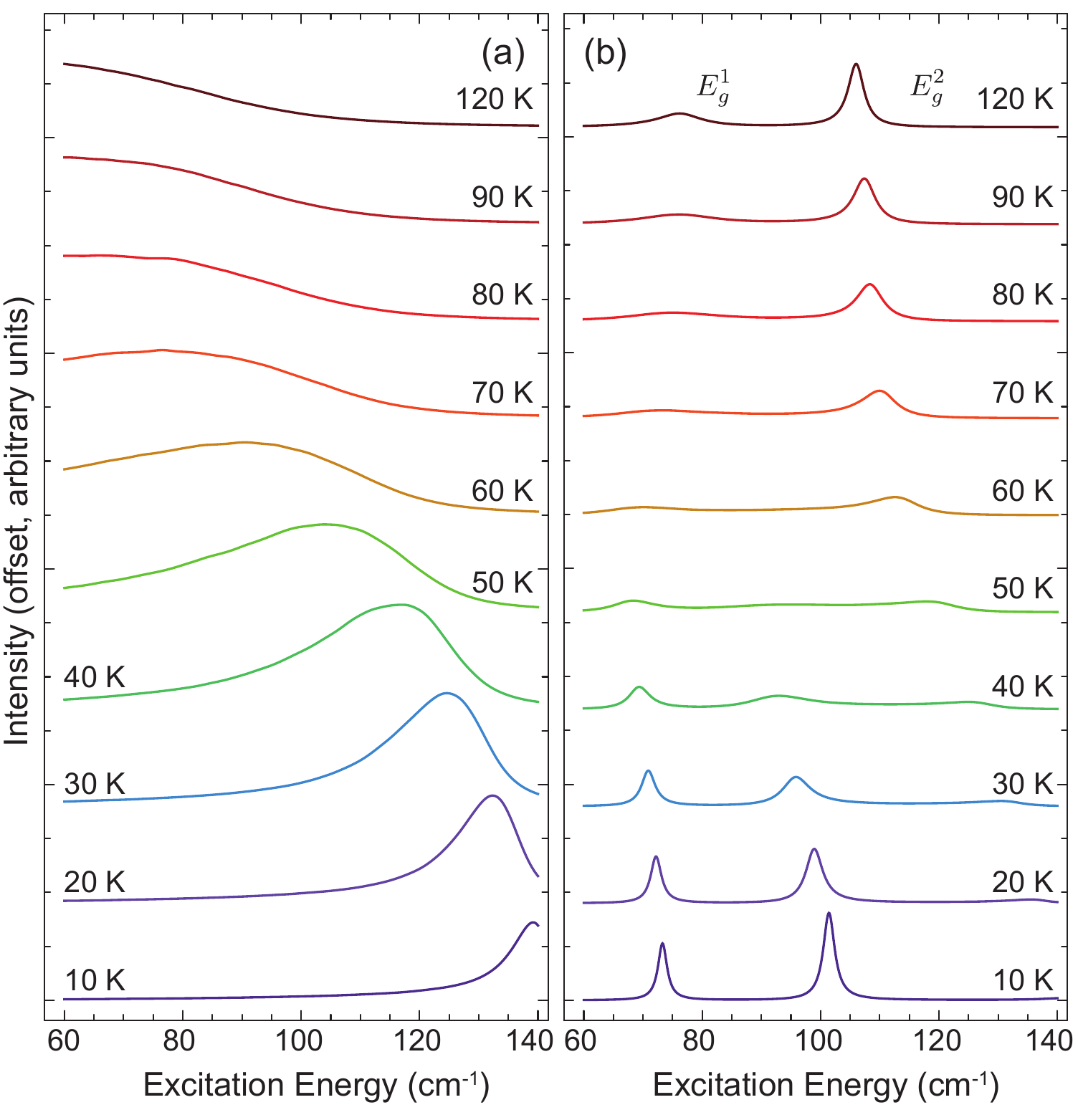}
\caption{(a) Computed 2-magnon Raman scattering intensity in the VV polarization (other polarizations are nearly identical) as a function of temperature. (b) Computed phonon Raman lineshapes (2-magnon intensity omitted).}
\label{fig:lineshape}
\end{figure}

In order to estimate the temperature-dependent bond-bond correlation functions appearing in equations (\ref{eqn:bond-bond1}) and (\ref{eqn:bond-bond2}), we employ classical Landau-Lifshitz (LL) spin dynamics, implemented in a modified version of the Sunny.jl package\cite{sunny,zhang2021classical,dahlbom2022geometric,dahlbom2022langevin}. In this approach, a distribution of initial spin states is sampled according to classical thermal equilibrium at a given temperature $T$. For each initial configuration, the LL equations of motion ($d\mathbf{S}_i/dt = (d\mathcal{H}/d\mathbf{S}_i) \times \mathbf{S}_i$) are iterated over a series of small time steps to generate trajectories $\mathbf{S}_i(t)$ in real time. From these trajectories, classical correlation functions of interest are computed as:
\begin{align}
C^R(\omega) = & \  \int_0^\infty dt \ e^{i\omega t}\langle \mathcal{R}(t) \mathcal{R}(0)\rangle_{\rm classical}
\\
    C_\nu^\Pi(\omega) = & \  \int_0^\infty dt \ e^{i\omega t}\langle \mathcal{A}_{v}(t) \mathcal{A}_{v}(0)\rangle_{\rm classical}
\end{align}
The two-magnon Raman intensity and imaginary part of the phonon self-energy are then approximated by:
\begin{align}
I_{\rm 2mag}(\omega) \propto & \ \frac{|\omega|}{k_BT}[1+n_B(\omega)]C^R(\omega) \label{eq:ramanfinal}
\\
    \Pi_{q,\nu}(i\nu_n) =& \  -\int_{-\infty}^{\infty}d\omega \ \frac{|\omega|}{k_BT}\frac{C_{q,\nu}^\Pi(\omega)}{i\nu_n-\omega}
\end{align}
It may be noted that the specific relationship between quantum and classical correlation functions is somewhat ambiguous; the prefactor of $\frac{|\omega|}{k_BT}[1+n_B(\omega)]$ in equation (\ref{eq:ramanfinal}) is the ``harmonic approximation'', familiar in molecular dynamics\cite{ramirez2004quantum}, and represents the standard approach for dynamical spin correlations in Sunny.jl. While the harmonic approximation does not precisely reproduce the expected quantum sum rules at elevated temperatures (see Ref.~\onlinecite{dahlbom2024quantum}), in the present case of $S = 5/2$, there exist relatively small discrepancies between the quantum and classical sum rules due to the large value of $S$. For the purpose of benchmarking the computed spin-phonon couplings, we find the LL dynamics to be suitably accurate. 
For each temperature, we perform dynamics calculations on 2D systems with $100 \times 100$ unit cells (200 sites), averaginf over 1000 initial spin states. We employ a time step of $\Delta t = 0.02/J_1$.  We use a total of 240 frequency points with a maximum frequency of 25 meV while calculating the intensity.

In Fig.~\ref{fig:lineshape}(a), we first show the computed temperature dependence of the magnetic Raman scattering response. As previously explained in Ref.~\onlinecite{mai2021magnon}, the magnon dispersion is relatively flat near the top of the band at $\sim 9$ meV, which results in a large peak in the two-magnon density of states at roughly twice this energy (18 meV $\approx$ 145 cm$^{-1}$). With increasing temperature, this peak broadens and softens continuously, eventually becoming quasi-elastic for $T>T_N$.

In Fig.~\ref{fig:selfenergy}(a-d), we show the computed temperature dependence of the $q = 0$ phonon self-energy for the $E_g^1$ and $E_g^2$ bands. The imaginary part of $\Pi_{Eg1}(\omega,T)$ and $\Pi_{Eg2}(\omega,T)$ have nearly identical form as the Raman scattering intensity, as all three quantities probe dynamical bond-bond correlations of similar form. The self-energy for the $E_g^2$ band is moderately larger than that of the $E_g^1$ band due to larger magnitude of the spin-phonon couplings $A_{(ij)}$. Very small wiggles in the computed self-energies, particularly at higher temperatures, are the result of numerical noise. 

In Fig.~\ref{fig:selfenergy}(e,f), we show the derived phonon lifetimes $\tau_\nu$ and approximate renormalized center frequency $\tilde{\omega}_\nu$, evaluated from the analytic continuation of the self-energy:
\begin{align}
    \tau_\nu \equiv & \ \frac{2\pi\hbar}{\text{Im}[\Pi_{0,\nu}(\omega_\nu^0+i\eta)]}
    \\
    \tilde{\omega}_\nu \equiv & \ \omega_\nu^0 - \text{Re}[\Pi_{0,\nu}(\omega_\nu^0+i\eta)]
\end{align}
where $\omega_\nu^0$ is the bare phonon frequency obtained from the {\it ab-initio} phonon calculations. Remarkably, we see a nearly quantitative agreement with the experiment of Ref.~\onlinecite{mai2021magnon}. For the $E_g^1$ phonons, the computed minimum lifetime is $\tau_{\rm min} \sim 4.6$ ps, which occurs at $T\sim 65$ K. Experimentally, this minimum was estimated to be $\sim 5$ ps, occurring at $\sim 70$ K. This temperature coincides with the main peak of Im[$\Pi_{Eg1}(\omega)$] passing the bare phonon frequency. In both experiment and our calculations, the lifetime is weakly temperature dependent above the minimum temperature. For the $E_g^2$ phonons, the computed lifetime minimum occurs instead at $\sim 50$ K, with a value of $\tau_{\rm min} \sim 2.1$ ps. This coincides with the experimentally estimated minimum of $\sim 4$ ps, also at $50$ K. In both experiment and our calculations, the lifetime increases rapidly with temperature above 50 K. 

The computed shifting of the center frequency (Fig.~\ref{fig:selfenergy}(f)) reproduces well the overall trends observed in experiment, although we predict somewhat larger shifts at intermediate temperatures than reported in Ref.~\onlinecite{mai2021magnon}. On this point, it may be noted that the broad, multi-peaked experimental lineshapes with both phononic and magnetic contributions can make precise definition of the phonon peak position somewhat ambiguous. We also do not account theoretically for regular magnetostriction effects, i.e.~the shifting of phonons due to changes in the crystal geometry resulting from the temperature dependence of the magnetic order parameter. Such effects are associated mostly with the $\mathbb{K}$ spin-phonon couplings, which reflect spin-dependent changes to the atomic forces. Experimentally, all optical phonons are observed to shift moderately with temperature due to this effect\cite{gillard2024spin}, which is independent of the two magnon-phonon avoided crossing effect included here. It was noted by the authors of Ref.~\onlinecite{mai2021magnon} that their experimental fits could be slightly improved by including additional temperature-dependent shifts of the phonons. We take this to imply an additional effect of magnetostriction, not included here.

For the purpose of comparison with experiment, in Fig.~\ref{fig:lineshape}(b), we show the predicted temperature evolution of the phonon lineshapes, according to equation (\ref{eq:phononlineshape}). For the bare phonon frequencies $\omega_\nu^0$, we use those obtained from the {\it ab-initio} phonon calculations. We take the bare width to be $\Gamma = 0.1$ meV, which is consistent with the experimental resolution in Ref.~\onlinecite{mai2021magnon}. We also take the phonon oscillator strengths to satisfy $|R_{Eg2}|^2 = 2|R_{Eg1}|^2$, in order to match the appearance of the experimental spectrum. Within the magnetically ordered state ($T < T_N = 74$ K), with increasing temperature, the $E_g^1$ phonons broaden significantly, as the maximum of the two-magnon DOS descends to lower energy. For $T>T_N$, the $E_g^1$ phonons remain broad, as they fall well within the energy range of the quasielastic tail of the incoherent spin dynamics in the paramagnetic phase. For the higher energy $E_g^2$ phonons, the maximum of the two-magnon DOS passes the phonon energy at $T\sim 50$ K. Around this temperature, the phonon lineshape develops a very broad double-peaked form reminiscent of an avoided two-magnon/phonon crossing. Above 50 K, the peak on the higher energy side evolves into the main phonon peak, and the width of the resonance continuously decreases, until the low-temperature linewidth is recovered. Above $T_N$, these phonons lie at the energetic edge of the quasi-elastic spin response, such that they are mostly energetically decoupled from the spin fluctuations.

Altogether, these findings agree with and support the findings of Ref.~\onlinecite{mai2021magnon}. The remarkable agreement between the computed values and experiment should lend significant confidence to the above described {\it ab-initio} spin-phonon coupling approach.

\section{Conclusions and Outlook}\label{sec:conclusions}
In this work, we have motivated the need to consider spin-phonon scattering in various contexts. The form of the spin-phonon coupling was discussed with reference to generic multipolar operators, and motivated by a simple model using analytical perturbation theory. Importantly, the full spin-phonon Hamiltonian contains terms that couple spin degrees of freedom with atomic displacement and atomic momenta. The appearance of the momentum couplings (such as $\mathbf{S}\cdot(\mathbf{u}\times\mathbf{p})$) was shown to be related to structural dependence of the specific spin-orbital composition of the ground state moments. Given that such terms are implicated in the controversial phonon Hall effect, their understanding is crucial. While we leave detailed discussion of these terms in specific materials for future works, here we have introduced a generic and relatively inexpensive {\it ab-initio} scheme for computing arbitrary multipole-phonon couplings with full $q$-dependence. This approach overcomes various deficiencies of more traditional DFT frozen phonon total energy approaches, including: (i) Providing estimates for the momentum couplings through inclusion of the phonons as genuine dynamical variables, and (ii) Including the full many-body structure of the local magnetic degrees of freedom, which is important for capturing their precise spin-orbital structure. We note that very recent developments\cite{bonini2023frequency,ren2024adiabatic} have also provided DFT-based approaches for estimating generic spin dipole-phonon couplings. It will be interesting to benchmark the current approach with these novel methods in the future. 

In the present manuscript, we benchmarked the described method for the layered honeycomb material MnPSe$_3$, which (by choice) does not have strong spin-orbit effects, leading to a relatively simple magnetic Hamiltonian and form of the spin-phonon coupling. Combining the computed spin-phonon couplings with classical Landau-Lifshitz dynamics to estimate the dynamical bond-bond correlations allows for the evaluation of the temperature and frequency dependent phonon self-energy. This approach was shown to capture many aspects of the experimental response, including (i) distinguishing phonon modes that couple strongly or weakly with the spins, and (ii) reproducing the temperature-dependent, and band-dependent $q =0$ optical phonon lifetimes reported in Ref.~\onlinecite{mai2021magnon}. These successes validate both the computed couplings and dynamical bond-bond correlations. Altogether, this approach allows for the estimation of general $q$-, and $T$-dependent spin-phonon scattering rates, which are also relevant for thermal transport\cite{hess2007heat,chernyshev2015thermal,chernyshev2016heat} and ultrasound attenuation\cite{zhou2011spinon,poirier2014ultrasonic,streib2015elastic,metavitsiadis2020phonon,feng2022sound}. 

At this point, one may ask whether the effort to implement spin-phonon couplings within a dCEH framework is truly necessary. Indeed, the material at focus in this work (MnPSe$_3$) could likely be well described by more traditional DFT approaches. However, this is not the case for many vdW materials and other quantum magnets of current interest. For example, the present authors recently showed that the isostructural compound FePS$_3$ has strongly spin-orbital coupled local moments\cite{dhakal2024hybrid}, and that the low-lying excitations should be viewed as complex mixtures of magnons and spin-orbital excitons. This material exhibits strong spin-phonon coupling as implied by avoided energetic crossings of phonons and spin excitations under magnetic field\cite{vaclavkova2021magnon,liu2021direct,zhang2021coherent}. FePS$_3$ has also been implicated as a possibly hosting topological spin/phonon hybridized modes\cite{to2023giant,cui2023chirality,luo2023evidence}. However, a key element for microscopic understanding of these effects is that the rich spin-orbital nature of the low-lying states must have consequences for the spin-phonon coupling; direct hybridization of one-magnon and one-phonon excitations is otherwise unlikely because a naive coupling like $\mathbf{u} \cdot \mathbf{S}$ is forbidden due to being odd under time reversal. Consistently, MnPSe$_3$ does not form magnon-polarons\cite{liao2024spin}; the fact that the primary form of spin-phonon coupling is $A_{ij} \mathbf{S}_i \cdot \mathbf{S}_j$, and the magnetic order is collinear implies a lack of mixing of one-magnon and one-phonon modes. Various phenomenological couplings have been proposed for materials like FePS$_3$ that do form magnon-polarons, which include noncolinear couplings in the intersite exchange or single-ion anisotropy terms, but these currently lack a microscopic basis. The same holds for many of the other materials mentioned in the introduction that exhibit signatures of spin-phonon coupling in their transport and excitations. 

An important corollary of this work is that one can use the temperature-dependent properties of the phonons as internal probes of spin correlation functions, provided the form of the spin-phonon coupling is known. In the present example, the magnetic Raman scattering and $q=0$ phonon lifetimes probe essentially the same two-magnon excitations, such that the phonons themselves do not provide access to additional correlation functions. However, measurements of phonon lifetimes at finite momentum, for example via ultrasound attenuation, can provide access to low-energy bond-bond correlations at finite momentum, which would be otherwise difficult to probe. Here, robust first principles spin-phonon methods are of significant utility for analyzing such experiments.

In summary, the goal of this work has been to demonstrate an {\it ab-initio} approach capable of addressing a myriad of questions regarding spin-phonon coupling, hybrid spin-phonon excitations, and thermal transport including in materials with strong spin-orbit coupling. We look forward to pursuing these phenomena in future studies.

\begin{acknowledgments}
The authors would like to thank M. Ozerov for many inspiring discussions on MnPSe$_3$ and other materials. We further thank A. Hight Walker and K. F. Garrity for detailed insights into their modeling of data on MnPSe$_3$, as well as S. Ullah, S. Biswas, and N. A. W. Holzwarth for assistance with ab-initio phonon calculations. We also acknowledge discussions with M. Mourigal and particularly S. Quinn regarding Sunny.jl. Finally, we are thankful for discussions with S. Ren and D. Vanderbilt regarding spin-phonon coupling. This research was funded by the Center for Functional Materials at WFU through a pilot grant, and Oak Ridge Associated Universities (ORAU) through the Ralph E. Powe Junior Faculty Enhancement Award to S.M.W. Computations were performed using the Wake Forest University (WFU) High Performance Computing
Facility \cite{WakeHPC}, a centrally managed computational resource available to WFU researchers including faculty, staff, students, and collaborators.
\end{acknowledgments}

\appendix

\bibliographystyle{apsrev}
\bibliography{mnps3}

\section{Intersite Coulomb Tensors} \label{sec:appendix}
Here, we give the estimated intersite Hund's coupling tensors, which represent the downfolded ferromagnetic Goodenough-Kanamori exchange in the $d$-orbital Wannier functions discussed in Section \ref{sec:mnpse3_abinitio_treatment}. These are written with respect to cubic coordinates, in which $x+y+z \ || \ c^*$, and $x$, $y$, and $z$ are otherwise aligned with the Mn-Se bond axes. For the nearest neighbor Mn-Mn bond in which the cubic $z$-axis is perpendicular to the plane of the edge-sharing bond [denoted by $J_{1a}$ in Fig.~\ref{fig:phonons}(g)], we use:
\begin{align}
    J_{H,ij}^{\alpha\beta} = \left(\begin{array}{r|ccccc}
    &d_{xy}&d_{xz}&d_{yz}&d_{z^2}&d_{x^2-y^2}\\
    \hline
    d_{xy}& 0.51 & 0.10 & 0.11 &  0.71 & 1.64\\
    d_{xz}& 0.10 & 0.05 & 0.09 & 0.24 &  0.19\\
    d_{yz}& 0.11 & 0.09 &  0.05 & 0.23 & 0.19\\
    d_{z^2}& 0.71 & 0.24 & 0.23 &  1.00& 0.83\\
    d_{x^2-y^2} & 1.64 & 0.19 & 0.19 & 0.83 &  1.80
    \end{array} \right)
\end{align}
(in units of meV). For the high-spin $d^5$ case, all orbitals are equally occupied, such that the impact of these intersite Coulomb interactions on the final magnetic couplings can be estimated via $J_{FM} = \frac{1}{2S^2}\sum_{\alpha,\beta} J_{H,ij}^{\alpha\beta} = -0.96$ meV. That is to say, we estimate the effect of the ferromagnetic ligand exchange contributions on the nearest neighbor exchange to be roughly half the antiferromagnetic exchange contribution.  It may be noted that this matrix is dominated by the $e_g$ orbitals, which hybridize most strongly with the ligand $p$-orbitals. For the second neighbor bond perpendicular to this bond, we have:
\begin{align}
    J_{H,ij}^{\alpha\beta} = \left(\begin{array}{r|ccccc}
    &d_{xy}&d_{xz}&d_{yz}&d_{z^2}&d_{x^2-y^2}\\
    \hline
    d_{xy}& 0.  & 0.04& 0.06& 0.29& 0.01\\
    d_{xz}& 0.01& 0.02& 0.02& 0.09& 0.01\\
    d_{yz}& 0.02& 0.01& 0.01& 0.02& 0.01\\
    d_{z^2}& 0.08& 0.04& 0.06& 0.23& 0.03\\
    d_{x^2-y^2} & 0.01& 0.09& 0.07&  0.37& 0.02
    \end{array} \right)
\end{align}
Here, the effective Coulomb interactions are significantly smaller. $J_{FM} = -0.13$ meV. This is sufficient to nearly completely compensate the second neighbor antiferromagnetic contribution to the exchange. Finally, For the third neighbor bond denoted by $J_{3a}$ in Fig.~\ref{fig:phonons}(g), we use:
\begin{align}
    J_{H,ij}^{\alpha\beta} = \left(\begin{array}{r|ccccc}
    &d_{xy}&d_{xz}&d_{yz}&d_{z^2}&d_{x^2-y^2}\\
    \hline
    d_{xy}& 0.05 & 0.03 & 0.04 & 0.1 &  0.31\\
    d_{xz}& 0.03 &  0.  & 0.04 &  0.07 &  0.13\\
    d_{yz}& 0.04 & 0.04 & 0.  &  0.08 & 0.12\\
    d_{z^2}& 0.1 & 0.07 & 0.08 & 0.16 &  0.47\\
    d_{x^2-y^2} & 0.31 & 0.13 & 0.12 & 0.47 &  1.22
    \end{array} \right)
\end{align}
which leads to $J_{FM} = -0.34$ meV. Thus, similar to the first neighbor couplings, we estimate the ferromagnetic correction is roughly half of the antiferromagnetic contribution. As noted in the main text, the structural dependence of these matrices is weak compared to the modulation of the direct $d-d$ hopping between nearest neighbors, such that they are not a significant source of spin-phonon coupling.

\end{document}